\documentclass[pre,aps,onecolumn,superscriptaddress,nofootinbib]{revtex4-1}
\pdfoutput=1
\usepackage{amsmath, amsthm, amssymb}
\usepackage{amsfonts}
\usepackage{graphicx}
\usepackage{dcolumn}
\usepackage{bm}
\usepackage{textcomp}
\usepackage[normalem]{ulem}

\usepackage[titletoc,title]{appendix}

\usepackage{epsfig}
\usepackage{subfigure}
\usepackage{url}
\usepackage{float}
\usepackage{color}
\usepackage{cases}

\usepackage{colordvi}
\usepackage[usenames,dvipsnames]{xcolor}

\usepackage[colorlinks=true, urlcolor=blue, anchorcolor=blue, citecolor=blue,filecolor=blue,linkcolor=blue,menucolor=blue
]{hyperref}

\usepackage[titletoc,title]{appendix}

\usepackage{color}

\begin{document}
\title{Statistics of first-passage Brownian functionals}

\author{Satya N. Majumdar}
\affiliation{LPTMS,  CNRS,  Univ.   Paris-Sud,  Universit\'{e}  Paris-Saclay,  91405  Orsay,  France}
\email{satya.majumdar@u-psud.fr}
\author{Baruch Meerson}
\affiliation{Racah Institute of Physics, Hebrew University of
Jerusalem, Jerusalem 91904, Israel}
\email{meerson@mail.huji.ac.il}

\begin{abstract}

We study the distribution of first-passage functionals of the type ${\cal A}= \int_0^{t_f} x^n(t)\, dt$ where
$x(t)$ represents a Brownian motion (with or without drift) with diffusion constant $D$, starting at $x_0>0$,
and $t_f$ is the first-passage time to the origin. In the driftless case, we compute exactly, for all $n>-2$, the probability density
$P_n(A|x_0)=\text{Prob}.(\mathcal{A}=A)$. We show that $P_n(A|x_0)$ has an essential singular tail as
$A\to 0$ and a power-law tail $\sim A^{-(n+3)/(n+2)}$ as $A\to \infty$. The leading essential singular
behavior for small $A$ can be obtained using the optimal fluctuation method (OFM), which also predicts
the optimal paths of the conditioned process in this limit. For the case with a drift toward the origin, where no exact
solution is known for general $n>-1$, we show that the OFM successfully predicts the tails of the distribution.
For $A\to 0$ it predicts the same essential singular tail as in the driftless case. For $A\to \infty$ it predicts
a stretched exponential tail $-\ln P_n(A|x_0)\sim A^{1/(n+1)}$ for all $n>0$. In the limit of large
P\'eclet number $\text{Pe}= \mu x_0/(2D)$, where $\mu$ is the drift velocity toward the origin, the OFM predicts an exact large-deviation scaling behavior, valid for all $A$:
$-\ln P_n(A|x_0)\simeq\text{Pe}\, \Phi_n\left(z= A/\bar{A}\right)$, where $\bar{A}=x_0^{n+1}/{\mu(n+1)}$ is the mean value of $\mathcal{A}$ in this limit.
We compute the rate function $\Phi_n(z)$ analytically for all $n>-1$. We show that, while for $n>0$
the rate function $\Phi_n(z)$ is analytic for all $z$,
it has a non-analytic behavior at $z=1$ for $-1<n<0$ which can be interpreted as a dynamical phase transition. The order of
this transition is $2$ for $-1/2<n<0$, while for $-1<n<-1/2$ the order of transition is $1/(n+1)$; it changes continuously with $n$.
We also provide an  illuminating  alternative derivation of the OFM result by using a WKB-type asymptotic perturbation theory for
large $\text{Pe}$. Finally, we employ the OFM to study the case of $\mu<0$ (drift away from the origin). We show that, when the process is conditioned on reaching the origin, the distribution of $\mathcal{A}$ coincides with the distribution of $\mathcal{A}$ for $\mu>0$ with the same $|\mu|$.

\end{abstract}

\maketitle

\nopagebreak

%
%
%
%
%
%
%

\section{Introduction}
\label{intro}

Functionals of Brownian motion appear naturally in many different contexts
spanning across physics, chemistry, biology, computer science and mathematics (see \cite{BF2005} for a review).
Statistical properties of the functionals of a one-dimensional Brownian motion over a fixed time interval
have been studied extensively since the celebrated Feynman-Kac formula~\cite{K1949}.
Another class of
functionals of one-dimensional Brownian motion have also attracted quite a lot of attention, namely the
\emph{first-passage} Brownian functional, defined up to the time
of first passage of the
Brownian motion, starting say at $x_0>0$, to a certain point in space, \textit{e.g.}
the origin~\cite{BF2005}. More precisely, let us consider
a one-dimensional Brownian motion $x(t)$ with diffusion constant $D$ that starts at $x_0>0$ and
evolves in time via
the Langevin equation
\begin{equation}
\frac{dx}{dt}= \eta(t)\, ,
\label{bm.1}
\end{equation}
where $\eta(t)$ is the Gaussian white noise with zero mean and the correlator $\langle \eta(t)\eta(t')\rangle=
2\,D\, \delta(t-t')$.
Let $t_f$ denote the first time the process $x(t)$ crosses the origin. Clearly $t_f$ is a
random variable that varies from trajectory to trajectory. Let us define a random functional
\begin{equation}
\mathcal{A}= \int_0^{t_f} U(x(t))\, dt\,,
\label{func.1}
\end{equation}
where $U(x)$ can, in principle, be an arbitrary function. One is interested in computing the probability
distribution $P(A|x_0)$ that the functional $\mathcal{A}$ takes a specified value $A$, given the starting position of the particle $x_0$.  Motivated by several physical examples (see below), we will focus here on scale-free functionals, where
$U(x)=x^n$, with $n>-2$ as we justify later. Thus our object of interest is the probability density function (PDF) $P_n(A|x_0)$ that the first-passage functional
\begin{equation}
\mathcal{A}= \int_0^{t_f} [x(t)]^n\, dt
\label{funcxn.1}
\end{equation}
takes the value $A$. There are many examples where this family of functionals is of relevance.
For example,  $n=0$ corresponds to $\mathcal{A}=t_f$, that is the first-passage time itself, whose exact distribution is well known~\cite{Rednerbook,BMS2013}:
\begin{equation}
P_0(t_f|x_0)= \frac{x_0}{\sqrt{4\,\pi\,D}}\,t_f^{-3/2}\, \exp\left(-\frac{x_0^2}{4\,D\, t_f}\right)\, .
\label{fpdist.1}
\end{equation}
For large $t_f\gg x_0^2$, the PDF $P_0(t_f|x_0)$ has a power-law tail, $P_0(t_f|x_0)\sim
t_f^{-3/2}$. In contrast, at $t_f\to 0$, the PDF has an essential singularity $P_0(t_f|x_0)\sim
\exp\left(-\frac{x_0^2}{4\,D\, t_f}\right)$. Another example concerns the case $n=1$, where $\mathcal{A}=\int_0^{t_f} x(t) dt$
is the area swept by the Brownian motion till its first-passage time,  and its distribution
was computed exactly in Ref.~\cite{KM2005}. This particular case $n=1$ has many applications ranging
from queueing theory and combinatorics, all the way up to the statistics of avalanches in self-organized
criticality~\cite{KM2005}. For example, in the context of the queueing theory, $x(t)$ may represent
the length of a queue in front of a ticket counter during the so called `busy period' (say in the morning) and
$\mathcal A$ represents the total serving time of the customers during the busy period.
The same functional $\mathcal A$ also appears in the study of the distribution of avalanche
sizes in the directed Abelian sandpile model~\cite{DR1989,K2004},
of the area of staircase polygons in compact directed percolation~\cite{PB1995,C2002,K2004} and
of the collapse time of a ball bouncing on a noisy platform~\cite{MK2007}.
The case $n=-3/2$
appears in an interesting problem of estimating the distribution of the lifetime of a comet in the solar system
(see \textit{e.g.} Ref.~\cite{H1961} and the discussion in Ref.~\cite{BF2005}). The case $n=-1/2$ appears in the context of computing
the distribution of the period of oscillation of an undamped particle in a random potential,
such as the Sinai potential~\cite{DM2001}.

Given the multitude of applications for different choices of $n$, it is natural to ask whether one can compute the
distribution $P_n(A|x_0)$ for arbitrary $n$. Our first main result in this paper is an exact solution
for $P_n(A|x_0)$ for arbitrary $n>-2$, for which $P_n(A|x_0)$ is well behaved.   As we show, $P_n(A|x_0)$  is given by the formula
\begin{equation}
P_n(A|x_0)= \frac{1}{\Gamma(\nu)}\, \left(\frac{\nu^2}{D}\right)^{\nu}\,
\frac{x_0}{A^{\nu+1}}\, \exp\left(-\frac{\nu^2  x_0^{1/\nu}}{D\,A}\right)\, , \quad\,\, \nu= \frac{1}{n+2}\, ,
\label{final_sol.0}
\end{equation}
where $\Gamma(\dots)$ is the gamma function. One can check that Eq.~(\ref{final_sol.0}) perfectly agrees with all the known solutions for $n=0$, $n=1$, $n=-1/2$ and $n=-3/2$. As one can see from Eq.~(\ref{final_sol.0}), a power-law tail at large $A$ and an essential singular behavior at $A\to 0$ appear for all admissible $n$.
In order to shed light on the nature of the essential singularity, we employ the optimal fluctuation method (OFM). The OFM has been successfully applied recently in several other problems, dealing with Brownian motion pushed to a large deviation regime by constraints ~\cite{GF,Meerson2019,SmithMeerson2019,MeersonSmith2019,Agranovetal2019}. Here we show that the OFM reproduces the essential singularity exactly by identifying the optimal, or most likely, path -- a special trajectory of the Brownian motion that makes a dominant contribution to the PDF $P_n(A|x_0)$  at $A \to 0$. In particular, we find that, when $A \to 0$, the most likely value of the first-passage time $t_f$ is finite for $-2<n<2$ and infinite for $n\geq 2$.

In the second part of this paper we study the same class of functionals as in Eq. (\ref{funcxn.1}), but for
a Brownian motion with a nonzero constant drift,
\begin{equation}
\frac{dx}{dt}= - \mu + \eta(t)\, ,
\label{drift_bm.1}
\end{equation}
starting at $x_0>0$. In Eq.~(\ref{drift_bm.1}),
$\eta(t)$ is the same zero mean, delta-correlated Gaussian white noise as before, and $\mu$ describes the
drift. If $\mu>0$, the Brownian particle drifts toward the origin, while for $\mu<0$ it drifts  away from the origin. The relative magnitude of the drift and diffusion is described by the dimensionless P\'{e}clet number,
\begin{equation}\label{pecletnumber}
\text{Pe} = \frac{\mu x_0}{2D}.
\end{equation}
For $\mu>0$ (drift toward the origin), the particle will surely cross the origin for the first time at
a finite $t_f$, and
we are again interested in the PDF $P_n(A|x_0)$  of the values $A$ of the functional (\ref{funcxn.1}) for general $n$. It turns out that, for $\mu>0$,
this PDF is well behaved only for $n>-1$, as we explain below. For the case of $\mu<0$, the Brownian
particle escapes to infinity with a finite probability~\cite{Rednerbook}. We briefly consider the case of $\mu<0$ by conditioning the process on the non-escape.

For $\mu>0$ the PDF  $P_n(A|x_0)$  of the values of the functional~(\ref{funcxn.1}) was studied previously only for $n=0$, where  $\mathcal{A}$ is the mean passage time itself \cite{Rednerbook}, and $n=1$, where $\mathcal{A}$ represents the area under the drifted Brownian motion~\cite{KM2005,MK2007}. The case of $n=0$ is exactly solvable and well known \cite{Rednerbook}.
In the case of $n=1$ it was possible to obtain an exact solution for the Laplace transform of
$P_1(A|x_0)$ in terms of Airy functions~\cite{KM2005,MK2007}.
However, inverting this Laplace transform turned out to be extremely hard, and even extracting the asymptotic behavior
of $P_1(A|x_0)$ was quite nontrivial and technical, in particular for large $A$~\cite{KM2005}.
Several papers have been devoted to extracting the large $A$ asymptotics and
computing the moments of $A$ for $n=1$~\cite{KMM2007,KM2014}.
For other $n>-1$ and $\mu>0$ there are no known exact solutions (not even in the Laplace space).  This is an ideal situation
where one can apply the OFM and obtain powerful new results. Here we take this course and obtain the exact leading-order asymptotic behaviors of $P_n(A|x_0)$ both at $A\to 0$ and $A\to \infty$, for arbitrary $n>-1$. In particular, we show that for $n\geq 0$ these asymptotics are given by the expressions
\begin{eqnarray}
-\ln P_n(A|x_0)\simeq \begin{cases}
& \frac{1}{D} \frac{x_0^{n+2}}{(n+2)^2}\, \frac{1}{A},
\quad \quad A\to 0, \\
\\
& \frac{\beta_n}{2D}\, \mu^{\frac{n+2}{n+1}}\,
A^{\frac{1}{n+1}},\quad A\to \infty,
\end{cases}
\label{asymp.0}
\end{eqnarray}
where
\begin{equation}
\beta_n = (n+1) \left[\frac{\sqrt{\pi}\, 2^{1-1/n}}{n+2}\,
\frac{\Gamma(1+1/n)}{\Gamma(1/2+1/n)}\right]^{\frac{n}{n+1}}\, .
\label{beta_n.0}
\end{equation}
For $n=0$ Eqs.~(\ref{asymp.0})
agree with the asymptotics extracted from
the exact expression for $P_0(A|x_0)$ using the fact that $\lim_{n\to 0} \beta_n=1/2$ \cite{Rednerbook}. For $n=1$, they agree with those of
Refs.~\cite{KM2005,KMM2007}. The $A\to 0$ asymptotic in Eq. (\ref{asymp.0})
is valid for all $n>-1$,
is independent of $\mu$ and agrees with the
leading small-$A$ behavior, Eq.~(\ref{final_sol.0}), obtained for $\mu=0$.
The large-$A$ asymptotic in Eq. (\ref{asymp.0}) is valid only for $n\ge 0$.
For $-1<n<0$, the large-$A$ tail of $P_n(A|x_0)$ presumably exhibits a power-law behavior which is beyond the accuracy of the OFM.
The asymptotics~(\ref{asymp.0}) constitute the
second main result of our paper.

In the weak noise limit,
which corresponds to $\text{Pe}\to \infty$, see Eq.~(\ref{pecletnumber}),
the OFM becomes asymptotically exact for all $A$. We show that in this limit $-\ln P_N(A)$ exhibits a large deviation scaling of the form
\begin{equation}
-\ln P_n(A) = \text{Pe} \,\Phi_n\left(\frac{A}{\bar A}\right)\, ,
\label{scaling.0}
\end{equation}
where
\begin{equation}
\bar{A} = \frac{x_0^{n+1}}{\mu (n+1)}\, ,
\label{abar_def}
\end{equation}
and we compute analytically, for any $n>-1$, the rate function $\Phi_n(z)$.  When $n\geq 0$,
the rate function vanishes only at its unique minimum point $z=1$, that is at $A=\bar{A}$, where $\bar{A}$ is the mean value of $A$ in the limit of  $\text{Pe}\to \infty$. In this case $\Phi_n(z)$  has a quadratic behavior near the minimum point $z=1$, describing typical, Gaussian fluctuations of $A$. Furthermore, it diverges at $z\to 0$ and $z \to \infty$ leading to the asymptotic behaviors (\ref{asymp.0}).

For $-1<n<0$ the behavior of $\Phi_n(z)$ changes dramatically. At $z<1$ the rate
function $\Phi_n(z)$ continues to be nonzero, and its $z\to 0$ asymptotic corresponds
to the $A\to 0$ asymptotic of Eq.~(\ref{asymp.0}).  Remarkably, the rate function is
equal to zero at all $z\geq 1$, and we uncover a dynamical phase transition at $z=1$.
For $-1/2<n<0$ we obtain $\Phi_n(z)\sim (1-z)^2$ as $z\to 1$ from below, so
this dynamical phase transition is of the second order.
However, for
$-1<n<-1/2$, we find $\Phi_n(z)\sim (1-z)^{1/(n+1)}$ as $z\to 1$ from below. Here the order of transition continuously depends on $n$ and varies from $2$
at $n\to -1/2$ to infinity at $n\to -1$. The order of transition is thus in general
non-integer and even non-rational.  This dynamical phase transition and, more
generally, the exact rate function $\Phi_n(z)$ for all $n>-1$, alongside with the
predicted optimal paths of the Brownian motion, conditioned on a specified $A$,
constitute the third main result of this paper.

Finally, we employ the OFM to study the case of $\mu<0$ (drift away from the origin), when the process is conditioned on reaching the origin. Here we show that the distribution of $\mathcal{A}$ coincides with the distribution of $\mathcal{A}$ for $\mu>0$ with the same $|\mu|$.

The rest of the paper is organised as follows. In Section \ref{zerodrift} we consider the Brownian motion with zero drift.
We first obtain, in Section \ref{zerodriftexact}, $P_n(A|x_0)$ exactly for arbitrary $n>-2$. In Section \ref{OFMexplain} we show
how to obtain the small-$A$ asymptotics of $P_n(A|x_0)$, and explain the essential singularity, by using the OFM. In Section \ref{drift} we consider the Brownian motion with
a drift toward the origin $(\mu>0)$. We start with presenting some exact results in the particular cases $n=0$, $1$ and $2$. Then we show how one can apply the OFM  in order to compute, in the limit of $\text{Pe} \to \infty$,
the exact rate function $\Phi_n(z)$ for all $n>-1$.  In Sec. \ref{alternative} we reproduce our OFM results for $\Phi_n(z)$ by a different, albeit related method: applying a variant of WKB approximation to the exact equation for the Laplace transform of $P_n(A|x_0)$.   We conclude with a summary and discussion in Section \ref{discussion}. The case $\mu<0$ is  considered
in the Appendix.

\section{Brownian motion with zero drift}
\label{zerodrift}

\subsection{Exact Results}
\label{zerodriftexact}

Here we consider the Brownian motion with zero drift as described by Eq. (\ref{bm.1}).
In order to compute $P(A|x_0)$ for general $U(x)$, it is useful to consider its Laplace
transform
\begin{equation}
Q_p(x_0)= \left\langle e^{-p\, A}\right\rangle =\left\langle e^{-p\, \int_0^{t_f} U(x(t))\, dt}\right\rangle=
\int_0^{\infty} P(A|x_0)\, e^{-p\, A}\, dA \,.
\label{func.2}
\end{equation}
The angular brackets $\langle \dots\rangle$ denote averaging over all trajectories starting at $x_0$ (this
averaging includes averaging over the history as well as over $t_f$ itself). A nice property of this Laplace transform
is that one can derive a linear second-order ordinary differential equation (ODE) for
$Q_p(x_0)$ by treating the starting position $x_0$ as a variable. This is the ``backward" approach
since one varies the position at the initial time. For a simple derivation of this equation we refer the readers to
Ref.~\cite{BF2005} (see also ~\cite{MC2002,KM2005}).
The main idea, in words, is to evolve the trajectory from $x_0$ to a new starting position $x_0+dx_0$
in a small time interval $dt$ and then keep track of how $Q_p(x_0)$ evolves as a result. Skipping details, one obtains
\begin{equation}
D\, \frac{d^2 Q_p}{dx_0^2} - p\, U(x_0)\, Q_p(x_0)=0\, ,
\label{diff.1}
\end{equation}
valid for $x_0\ge 0$, with the boundary conditions:
\begin{equation}\label{BCs}
\text{(i)}\quad Q_p(x_0= 0)=1 \quad \text{and}\quad \text{(ii)}\quad Q_p(x_0\to \infty)\to 0\,.
\end{equation}
The condition (i) stems from the fact that in this case $t_f=0$ for a
well behaved $U(x)$, and hence $\langle e^{-p\, \int_0^{t_f} U(x(t))\, dt}\rangle \to 1$ as $x_0\to 0$. The condition (ii)
follows from the fact that, as $x_0\to \infty$, $t_f\to \infty$ as well. When $A$ is kept fixed, the resulting PDF $P(A|x_0)$, and its Laplace transform  $Q_p(x_0)$, must vanish.

Notice that Eq.~(\ref{diff.1}) is different
from the Feynman-Kac equation: the latter is a partial differential equation that involves time explicitly (since
it deals with functionals over a fixed time interval). In our case, since one sums over all possible
trajectories with different first-passage times, there is no explicit time dependence in the
equation for $Q_p(x_0)$.

Equation~(\ref{diff.1}) can be viewed as a Schr\"{o}dinger equation [with a bit unusual boundary condition (\ref{BCs})\text{(i)} for the ``wave function"] for a zero-energy particle, and solving it for arbitrary  potential $U(x_0)$ is not possible. Fortunately, for our choice  $U(x)= x^n$,
as in Eq. (\ref{funcxn.1}), the solution can be obtained in a closed form. Here
Eq.~(\ref{diff.1}) becomes
\begin{equation}
D\, \frac{d^2 Q_p}{dx_0^2} - p\, x_0^n\, Q_p(x_0)=0\, .
\label{diff.2}
\end{equation}
This equation has two linearly independent solutions
\begin{equation}\label{qplusminus}
q_1(x_0)=\sqrt{x_0}\, I_{\nu }\left(2\nu
   \sqrt{\frac{p}{D}}\,
   x_0^{\frac{1}{2\nu}}\right)\quad\text{and}\quad q_2(x_0)=\sqrt{x_0}\, K_{\nu }\left(2\nu
   \sqrt{\frac{p}{D}}\,
   x_0^{\frac{1}{2\nu}}\right),
\end{equation}
where $\nu=1/(n+2)$, $I_{\nu}(\dots)$ and $K_{\nu}(\dots)$  are the
modified Bessel functions of the first and second kind~\cite{AS,GR}, respectively, and
we assumed $n>-2$ strictly\footnote{For $n\leq -2$ and $p>0$, the only solution
of Eq. (\ref{diff.2}) that satisfies the
boundary condition $Q_p(x_0\to \infty)=0$ is
$Q_p(x_0)=0$ leading to $\mathcal{A}=\infty$.}.
As $x_0\to \infty$, $q_1(x_0)$ diverges, while  $q_2(x_0)$ tends to zero. Therefore, $q_1(x_0)$ should be discarded. The solution $q_2(x_0)$ is well-behaved at $x_0\to 0$. Normalizing it so as to obey the boundary condition (\ref{BCs}) \text{(i)},
we arrive at the desired Laplace transform
\begin{equation}
Q_p(x_0)= \frac{2 \nu^{\nu}}{\Gamma(\nu)\, D^{\nu/2}}\, p^{\nu/2}\, \sqrt{x_0}\,
K_{\nu}\left( 2\, \nu\, \sqrt{\frac{p}{D}}\, x_0^{\frac{1}{2\nu}}\right)\, , \quad\,\, \nu= \frac{1}{n+2} \,,\quad n>-2\,.
\label{lpsol.2}
\end{equation}
The inversion of this Laplace transform looks challenging, but we succeeded in performing it by  virtue of the following identity~\cite{GR}:
\begin{equation}
\int_0^{\infty} dz\, z^{-\nu-1}\, e^{-p\, z- \frac{\gamma}{z}}= 2\,
\left(\frac{p}{\gamma}\right)^{\nu/2}\, K_\nu\left( 2\, \sqrt{\gamma p}\right)\, .
\label{iden.1}
\end{equation}
Consequently, the Laplace inversion
\begin{equation}
{\cal L}_p^{-1}\left[ p^{\nu/2}\, K_\nu(2\sqrt{\gamma p})\right]= \frac{1}{2}\,
\gamma^{\nu/2}\, A^{-\nu-1}\, e^{-\gamma/A}\, .
\label{iden.2}
\end{equation}
Hence, using the identity (\ref{iden.2}) and choosing
$\gamma= (\nu^2/D)\, x_0^{1/\nu}$, one can invert Eq.~(\ref{lpsol.2}) and get an exact expression for our distribution $P_n(A|x_0)$,
valid for all $A>0$ and $x_0\ge 0$, once $n>-2$:
\begin{equation}
P_n(A|x_0)= \frac{1}{\Gamma(\nu)}\, \left(\frac{\nu^2}{D}\right)^{\nu}\,
\frac{x_0}{A^{\nu+1}}\, \exp\left(-\frac{\nu^2}{D\,A}\, x_0^{1/\nu}\right)\, , \quad\,\, \nu= \frac{1}{n+2}\, .
\label{final_sol.1}
\end{equation}
One can check that $P_n(A|x_0)$ is normalized to unity, $\int_0^{\infty} P_n(A|x_0)\, dA=1$.
Also, for $n=-3/2$, $-1/2$, $0$ and $1$, it reduces to the known results.
As one can see, the $A$-dependence of $P_n(A|x_0)$ in Eq. (\ref{final_sol.1})
is given by a product of just two factors: the power-law factor $A^{-\nu-1}$ that describes the large-$A$ decay
and the factor $\exp\left(- \frac{\nu^2 x_0^{1/\nu}}{D A}\right)$, which determines the much faster
small-$A$ decay and exhibits an essential singularity at $A\to 0$.  The power-law factor $A^{-\nu-1}$ can be obtained by the following scaling argument. Noting that for a Brownian motion one has $x(t)\sim t^{1/2}$ for large $t$, one obtains
for large $t_f$
\begin{equation}
{\cal A}= \int_0^{t_f} [x(t)]^n\, d t \sim  t_f^{(n+2)/2}\, .
\label{scaling.1}
\end{equation}
Now, the distribution of the first-passage time $t_f$ for large $t_f$ scales
as
$P_0(t_f|x_0)\sim x_0\,t_f^{-3/2}$, see Eq. (\ref{fpdist.1}).
Hence, using $P_0(t_f|x_0)\, dt_f = P_n(A|x_0)\, dA$
and plugging the scaling relation
into Eq. (\ref{scaling.1}), we obtain $P_n(A|x_0) \sim x_0\, A^{-\nu-1}= x_0\,
A^{-(n+3)/(n+2)}$ for $A\to \infty$\footnote{It follows from Eq.~(\ref{final_sol.1}) that the mean value of $\mathcal{A}$ is finite for $-2<n<-1$ and infinite for $n>-1$.}.

This scaling argument however fails to account for the
essentially-singular small-$A$ behavior, since one can no longer use the
scaling relation (\ref{scaling.1})
for small $A$. This large-deviation-type behavior,
however, is perfectly captured by the optimal fluctuation method
as we now demonstrate.

\subsection{Optimal fluctuation method explains essential singularity at $A\to 0$}
\label{OFMexplain}

When applied to the Brownian motion, the OFM essentially becomes geometrical optics
\cite{GF,Meerson2019,SmithMeerson2019,MeersonSmith2019,Agranovetal2019}. A natural starting point of the OFM is the probability of a
Brownian path $x\left(t\right)$, which is given, up to pre-exponential factors, by the Wiener's action, see
\textit{e.g.} Ref. \cite{BF2005}:
\begin{equation}\label{Action}
-\ln P=S=\frac{1}{4D}\int_{0}^{t_f} \dot{x}^2(t)\,dt .
\end{equation}
The distribution $P_n(A|x_0)$ can then be written as $P_n(A|x_0)= \left \langle \delta\left(A- \mathcal{A}\right)\right\rangle$ where, as in Eq.~(\ref{func.2}), the angular brackets denote an average over all Brownian trajectories, starting at $x_0$ and reaching
the origin for the first time at $t_f$, as well as over all possible values of $t_f$. The delta-function
can be replaced by its integral representation:
\begin{equation}
P_n(A|x_0)= \left \langle \delta\left(\mathcal{A}-A\right)\right\rangle
= \left\langle \frac{1}{2D}\int \frac{d\lambda}{2\pi i} \exp \left[ \frac{\lambda}{2D}
\left( \mathcal{A} -A\right)\right]
\right\rangle ,
\label{constraint.2}
\end{equation}
where the integration over $\lambda$ is along the vertical axis (the Bromwich contour) in the complex $\lambda$ plane.
This extra piece, added to the Wiener measure in Eq. (\ref{Action}), gives rise to an effective action functional
\begin{equation}
S_{\rm eff} = \frac{1}{2D}\left[ \frac{1}{2} \int_0^{t_f}\dot{x}^2(t)\,dt + \lambda
\left( \int_0^{t_f} [x(t)]^n dt -A\right)\right]\, .
\label{action_eff.1}
\end{equation}
Thus, one can interpret $\lambda$ as the Lagrange multiplier that enforces the constraint $\mathcal{A}=A$.
In the regime when the effective action $S_{\rm eff}$ is very large, the leading-order contribution to $P_n(A|x_0)$
can be obtained by the saddle point method. This requires minimizing $S_{\rm eff}$ from Eq.~(\ref{action_eff.1})
(i) over all trajectories $x\left(t\right)$ that start at $x_0$, satisfy the
condition  $x(t)>0$ for $0<t<t_f$, and arrive at $x=0$ at time $t_f$,
(ii) over all possible values of $t_f$, and (iii) over $\lambda$
so as to impose the constraint $\mathcal{A}=A$. It is convenient to think of $S_{\rm eff}$ as of the action of a Newtonian particle of unit mass with the time-independent Lagrangian
\begin{equation}
L_{\lambda} \left(x,\dot{x}\right)= \frac{\dot{x}^2}{2} + \lambda x^n,
\label{lagrangian.1}
\end{equation}
where the first terms describes the kinetic energy, and the second term corresponds to the effective potential
$V(x)=- \lambda\, x^n$. The extremal is described by the Euler-Lagrange equation
\begin{equation}\label{EL}
\ddot{x}(t)-\lambda n x^{n-1}(t)=0.
\end{equation}
Having solved this equation subject to all constraints, we will obtain the optimal (most likely) path of our
constrained Brownian motion.  The first integral of Eq.~(\ref{EL}) describes conservation of energy:
\begin{equation}
\frac{{\dot x}^2 }{2} - \lambda\, x^n = E\, ,\quad 0\leq t\leq t_f.
\label{energy.1}
\end{equation}
where the energy $E$ is a constant of motion. To determine $E$, we minimize $S_{\rm eff}$ in
Eq. (\ref{action_eff.1}) with
respect to $t_f$ at fixed $x_0$ and $\lambda$.  We have
\begin{equation}
\frac{dS_{\rm eff}}{dt_f}=\left(\frac{{\dot x}^2 }{2} + \lambda x^n\right) \Big|_{t=t_f}\, .
\label{energy.2}
\end{equation}
For $n>0$ $x^n (t_f)=0$, and the second term in the r.h.s of Eq.~(\ref{energy.2}) vanishes. Demanding that $dS_{\rm eff}/dt_f=0$, we obtain $\dot x (t_f)=0$. Here the optimal path is such that the particle stops at $t=t_f$, when it reaches $x=0$. Then, using Eq.~(\ref{energy.1}) at $t=t_f$, we obtain $E=0$.

For $n\leq 0$ the second term in the r.h.s of Eq.~(\ref{energy.2}) does not vanish at $t=t_f$ (for $n<0$ it diverges), and neither does the derivative $dS_{\rm eff}/dt_f$. Still, it can be shown that here too $S_{\rm eff}$ as a function of $t_f$ has its minimum which corresponds to $E=0$, although $S_{\rm eff}$ is not smooth at this point\footnote{For $n<0$ and $E>0$ the trajectory of the effective Newtonian particle in the potential $V(x)$ is unique and such that $\dot{x}<0$: the particle moves to the left. For $n<0$ and $E<0$ there are two possible solutions: one where the particle moves to the left, and the other where the particle first moves to the right, gets reflected from the potential $V(x)$ and then moves to the left until it reaches $x=0$ at $t=t_f$.}.

Plugging $E=0$ in  Eq.~(\ref{energy.1}), we obtain
\begin{equation}
\frac{{\dot x}^2}{2}  = \lambda\, x^n\, ,\quad 0\leq t\leq t_f.
\label{energy.4}
\end{equation}
By virtue of Eq.~(\ref{energy.4}), the effective action in Eq. (\ref{action_eff.1}), evaluated on the optimal path, is equal to
\begin{equation}
S_{\rm opt} = S_{\rm eff}\big|_{\rm optimal\,\, path}=\frac{\lambda A}{2D}\,.
\label{S_opt.1}
\end{equation}
To express $\lambda$ via $A$ and $x_0$, we have to integrate Eq. (\ref{energy.4}) and obtain the optimal path $x(t)$, satisfying the required constraints.  From Eq. (\ref{energy.4}) we obtain
\begin{equation}\label{firstorder}
\dot x= -\sqrt{2\lambda}\, x^{n/2}\,.
\end{equation}
The equation with the plus sign, $\dot x= \sqrt{2\lambda}\, x^{n/2}$,
must be discarded because it would drive the path to infinity and
lead to $\mathcal{A}=\infty$ for all $n$. Some of the further details
of the optimal path depend on whether $-2<n<2$, $n>2$, or
$n=2$, as we will now see.

\subsubsection{$n>2$}
\label{nmore2}

The solution of Eq.~(\ref{firstorder})
that obeys the initial condition
$x(0)=x_0$, can be written as
\begin{equation}\label{sol1_ngt2}
x(t)= \left[x_0^{1-\frac{n}{2}}+\frac{\sqrt{\lambda} (n-2)
   t}{\sqrt{2}}\right]^{-\frac{2}{n-2}}.
\end{equation}
Here the optimal path approaches zero only at $t \to \infty$. That is, the optimal value
of the first-passage time $t_f=\infty$.  The
constraint (\ref{funcxn.1}) (with time integration extended to infinity) yields
\begin{equation}\label{lambda_ngt2}
\lambda =\frac{2 x_0^{n+2}}{A^2 (n+2)^2},
\end{equation}
so the optimal path, for specified $x_0$ and $A$, is
\begin{equation}\label{optimalpath_ngt2}
x(t)= x_0 \left[1+\frac{(n-2)
   x_0^{n} t}{(n+2) A}\right]^{-\frac{2}{n-2}}.
\end{equation}
Plugging $\lambda$ from Eq. (\ref{lambda_ngt2}) into Eq.~(\ref{S_opt.1}), we get
\begin{equation}\label{S_opt_ngt2}
S_{\rm opt} = \frac{x_0^{n+2}}{(n+2)^2DA} \,.
\end{equation}
As a result,
\begin{equation}\label{PnOFM}
P_n(A|x_0)\sim \exp (-S_{\rm opt}) = \exp\left[-\frac{x_0^{n+2}}{(n+2)^2DA}\right]\,,
\end{equation}
and we must demand  $S_{\rm opt}\gg 1$ to justify the saddle point evaluation of the path integral. Equation~(\ref{PnOFM}) correctly reproduces the leading-order  singular behavior of the exact result
in Eq.~(\ref{final_sol.1}). This happens when $A\to 0$ at fixed $x_0$ and $D$, or for any $A$ when $D x_0^{-(n+2)} \to 0$.

\subsubsection{$-2<n<2$}
\label{nminus22}

Here Eq.~(\ref{sol1_ngt2}) continues to hold, but the optimal solution $x(t)$ has a
compact support $0\leq t<t_f$, where $t_f$ is a finite optimal first-passage time. In terms of $\lambda$
\begin{equation}
t_f = \frac{\sqrt{2}\, x_0^{1-\frac{n}{2}}}{\sqrt{\lambda} \,(2-n)}.
\label{tfl_nlt2}
\end{equation}
The constraint (\ref{funcxn.1}), with integration from $0$ to $t_f$, again yields Eq.~(\ref{lambda_ngt2}), Eq.~(\ref{optimalpath_ngt2}) (where $0\leq t\leq t_f$) and Eqs.~(\ref{S_opt_ngt2}) and~(\ref{PnOFM}), in full agreement
with the leading small-$A$ behavior of $P_n(A|x_0)$ in Eq.~(\ref{final_sol.1}). In terms of $A$ the optimal first-passage time (\ref{tfl_nlt2}) is
\begin{equation}
t_f= \frac{(n+2) A}{(2-n) x_0^{n}},\quad -2<n<2\,.
\label{tf_nlt2}
\end{equation}
In the particular case $n=1$ the optimal path~(\ref{optimalpath_ngt2})  is a parabola
\begin{equation}\label{nodriftparabola}
x(t)=\frac{x_0 (3A-x_0 t)^2}{9 A^2}, \quad 0\leq t\leq t_f= \frac{3A}{x_0}.
\end{equation}
The parabola is tangent to the $t$-axis at $t=t_f$.
For $n=0$ the optimal path~(\ref{optimalpath_ngt2}) is a straight line:
\begin{equation}\label{nodriftstraight}
x(t)=x_0\left(1-\frac{t}{A}\right), \quad 0\leq t\leq t_f\equiv A.
\end{equation}

\subsubsection{$n=2$}
\label{n2}

In the special case $n=2$, the solution of Eq.~(\ref{firstorder}), conditioned on $x_0$ and $A$, is
\begin{equation}
x(t)=x_0\, e^{-\frac{x_0^2 t}{2A}}\,,
\label{optimalpath_n2}
\end{equation}
so that the optimal first passage time is infinite. Here $\lambda= x_0^2/(8A^2)$, and the optimal action is described by Eq.~(\ref{S_opt_ngt2}) and, again, correctly describes the singular behavior of exact $P_2(A|x_0)$ from  Eq.~(\ref{S_opt_ngt2}) with $n=2$ at $A\to 0$.

\section{Brownian motion in the presence of drift}
\label{drift}

\subsection{Exact Results}
\label{driftexact}

Here we consider the Brownian motion in the presence of a nonzero drift $\mu$ which can be described by Eq.~(\ref{drift_bm.1}).
We are again interested in the PDF $P(A|x_0)$ of the first-passage functionals of the type $\mathcal{A}= \int_0^{t_f}U(x(t))\, dt$. Following the same line of arguments as in the driftless case~\cite{MC2002,KM2005}, one can obtain, for arbitrary $U(x)$,
a second-order ODE for the Laplace transform, $Q_p(x_0)= \int_0^{\infty} e^{-p\,A} P(A|x_0)\, dA$:
\begin{equation}
D\, \frac{d^2 Q_p}{dx_0^2} - \mu\, \frac{dQ_p}{dx_0}- p\, U(x_0)\, Q_p(x_0)=0\, .
\label{mu_diff.1}
\end{equation}
The boundary conditions (\ref{BCs}) continue to hold.
For our choice $U(x)= x^n$ Eq.~(\ref{mu_diff.1}) becomes
\begin{equation}
D\, \frac{d^2 Q_p}{dx_0^2} - \mu\, \frac{dQ_p}{dx_0}- p\, x_0^n\, Q_p(x_0)=0\, .
\label{mu_diff.2}
\end{equation}
Unlike for $\mu=0$, where exact solutions could be derived for arbitrary $n>-2$,
for $\mu>0$ we are aware of only three exactly solvable cases
for Eq.~(\ref{mu_diff.2}): $n=0$, $1$ and $2$, so let us briefly consider them.

\subsubsection{$n=0$}
\label{driftexact0}

Here $A= t_f$, and the solution of Eq. (\ref{mu_diff.2}) with
the boundary conditions~(\ref{BCs}) is elementary~\cite{MC2002,KM2005}:
\begin{equation}
Q_p(x_0)= \exp\left(\frac{\mu x_0}{2D} - \frac{x_0}{\sqrt{D}}\, \sqrt{\frac{\mu^2}{4D}+p}\right)
\label{n0_sol.1}
\end{equation}
This Laplace transform can be readily inverted to give the well-known exact distribution of the first-passage time \cite{Rednerbook}
\begin{equation}
P_0(t_f|x_0)= \frac{x_0}{\sqrt{4\,\pi\, D\, t_f^3}}\, \exp\left[- \frac{1}{4\,D\, t_f}\,(x_0- \mu\, t_f)^2\right]\, ,
\label{n0_sol.2}
\end{equation}
which is a simple extension of the driftless result~(\ref{fpdist.1}).

\subsubsection{$n=1$}
\label{driftexact1}

In this case, studied in Ref.~\cite{KM2005}, one obtains
\begin{equation}
Q_p(x_0)= \int_0^{\infty} P_1(A|x_0)\, e^{-p\, A}\, dA= \frac{e^{\frac{\mu x_0}{2D}}
   \text{Ai}\left(\frac{\mu ^2+4 p
   D  x_0}{4 p^{2/3}
   D^{4/3}}\right)}{\text{Ai}\left
   (\frac{\mu ^2}{4 p^{2/3}
   D^{4/3}}\right)}\,,
\label{n1_sol.1}
\end{equation}
where $\text{Ai}(z)$ is the Airy function. Inverting this Laplace transform exactly does not seem feasible.
Even extracting the asymptotic behaviors of $P_1(A|x_0)$ for large $A$ from this Laplace transform
is nontrivial. This was done
in Ref.~\cite{KM2005} by employing a rather technical method (see also ~\cite{KMM2007}). In contrast, the small-$A$
behavior is easy to derive as it effectively corresponds to the driftless case. The
asymptotic behaviors of $P_1(A|x_0)$ are given by~\cite{KM2005}
\begin{eqnarray}
P_1(A|x_0)\simeq \begin{cases}
& \frac{x_0}{\Gamma(1/3) (9D)^{1/3}}\, A^{-4/3}\,\exp\left(-\frac{x_0^3}{9\,D\,A}\right)\,,
\quad\quad A\to 0, \\
\\
& \left(\frac{2}{3}\right)^{1/4} \frac{x_0}{\sqrt{\pi}\, (2D)^{3/2}}\, \mu^{7/4}\, A^{-3/4}\, \exp\left(-\sqrt{\frac{2}{3}}\,
\frac{\mu^{3/2}}{D}\, \sqrt{A}\right)\,,   \quad\quad  A\to \infty \,,
\end{cases}
\label{n1_asymp.1}
\end{eqnarray}
where the leading-order $A\to 0$ asymptotic coincides with the corresponding asymptotic in Eq.~(\ref{asymp.0}) for $n=1$.

\subsubsection{$n=2$}
\label{driftexact2}

To our knowledge, this case has not been studied before. The general solution of Eq.~(\ref{mu_diff.2}) can be represented as a linear combination of two independent solutions, one of which decaying at infinity, and the other growing without limit. In view of the boundary condition~(\ref{BCs}) at $x_0\to \infty$, the growing solution must be discarded. The remaining arbitrary constant is chosen so that the solution obeys the boundary condition~(\ref{BCs}) at $x_0=0$.  Skipping details, we just present the solution for $Q_p(x_0)$:
\begin{equation}
Q_p(x_0) =  e^{\frac{\mu x_0}{2D}}\,
\frac{D_q\left[\left(\frac{4p}{D}\right)^{1/4}\, x_0\right]}{D_q(0)}
\,, \quad {\rm where}\quad\, q= -\frac{1}{2}- \frac{\mu^2}{8 D^{3/2} \sqrt{p}}\, ,
\label{n2_sol.1}
\end{equation}
and $D_q(z)$ is the parabolic cylinder function~\cite{AS}; note that  $D_q(0)= \sqrt{\pi}\, 2^{q/2}/\Gamma[(1-q)/2]$.
Inverting the Laplace transform (\ref{n2_sol.1}) exactly looks hopeless. Once again, while the
leading small-$A$ behavior coincides with the one in the driftless case, extracting the large $A$ asymptotics is not easy.

We are unaware of any other case except $n=0$, $1$ and $2$, when
Eq.~(\ref{mu_diff.2}) with the boundary conditions (\ref{BCs}) can be solved exactly.
That is, not even the Laplace transform can be determined exactly.
As we will now see, here comes the real power of the OFM. But before employing
the OFM, we present one more exact result: for the mean value of the random variable
${\cal A}=\int_0^{t_f} x^n(t)\, dt$ for any $n>-1$.

\vspace{0.4cm}

\noindent
{\bf Exact mean}. In terms of the Laplace transform $Q_p(x_0)$, the mean is given by
\begin{equation}
\langle {\cal A}\rangle\equiv m_1(x_0)= \Big \langle
\int_0^{t_f} x^n(t)\, dt\Big \rangle= - \frac{dQ_p(x_0)}{dp}\Big|_{p=0}\,,
\label{mean.1}
\end{equation}
where $Q_p(x_0)$ satisfies Eq. (\ref{mu_diff.2}). Taking the derivative of Eq. (\ref{mu_diff.2}) with respect to $p$ and
setting $p=0$, we obtain a simple differential
equation for $m_1(x_0)$:
\begin{equation}
D\, \frac{d^2 m_1(x_0)}{dx_0^2} - \mu\, \frac{dm_1(x_0)}{dx_0}= -x_0^n \, .
\label{mean_diff.1}
\end{equation}
It has to be solved subject to the following boundary conditions: (i) $m_1(x_0=0)=0$ and (ii) $m_1(x_0)$ cannot grow faster than a power law
as $x_0\to \infty$. The solution is straightforward, and we obtain, for $n>-1$,
\begin{equation}
m_1(x_0)= \frac{1}{D}\,
\int_0^{x_0} dx\, e^{ \mu x/D}\, \int_{x}^{\infty} y^n\, e^{-\mu y/D}\, dy\, .
\label{sol.1}
\end{equation}
Evaluating the double integral, we arrive at the exact result
\begin{eqnarray}
  m_1(x_0)&=& \frac{1}{\mu}\,\left\{ \frac{{x_0}^{n+1}}{n+1}+
\left(\frac{D}{\mu}\right)^{n+1}\,
\left[ e^{\mu x_0/D}\,
\Gamma\left(n+1, \frac{\mu\, x_0}{D}\right)- \Gamma(n+1)\right]\right\} \nonumber\\
 &=& \bar{A} \left\{1+\frac{1}{\text{Pe}^{n+1}}\frac{n+1}{2^{n+1}}\left[e^{2\text{Pe}} \Gamma\left(n+1,2 \text{Pe}\right)-\Gamma(n+1)\right]\right\}\,,\quad n>-1\,, \label{sol.2}
\end{eqnarray}
where $\Gamma(n+1,z)=\int_z^{\infty} e^{-u}\, u^n\, du$ is the
incomplete gamma function, and $\bar{A}$ is defined in Eq.~(\ref{abar_def}).  Note that, as $\text{Pe}$ tends to infinity, the function $e^{2\text{Pe}} \Gamma\left(n+1,2 \text{Pe}\right)$ behaves as $2^n \text{Pe}^n$. As a result, the exact mean value $m_1(x_0)$ approaches $\bar{A}$ from Eq.~(\ref{abar_def}) as $1/\text{Pe}$ when $\text{Pe} \to \infty$ \footnote{As one can check from Eq.~(\ref{sol.2}), $m_1(x_0)$ diverges in the driftless case $\mu=0$, or $\text{Pe}=0$, in agreement with our driftless result of Sec.~\ref{zerodriftexact} for $n>-1$.}.

By taking higher derivatives of $Q_p(x_0)$ with respect to $p$ and setting
$p=0$ one can derive differential equations for higher moments
and, in principle, solve them recursively. However,  this recursive procedure,
beyond the first moment, quickly becomes complicated. In addition, it does not shed light
on the tails of the distribution $P_n(A|x_0)$.  We will show now
how to obtain the distribution tails by using the OFM.

\subsection{Optimal fluctuation method}
\label{driftOFM}

By virtue of the Langevin equation~(\ref{drift_bm.1}), the probability of an unconstrained path $x\left(t\right)$ is now given by
\begin{equation}\label{Actiondrift}
-\ln P=S=\frac{1}{4D}\int_{0}^{t_f} \left(\dot{x}+\mu\right)^2\,dt \, .
\end{equation}
As in the driftless case in Eq. (\ref{action_eff.1}), taking into acount the constraint
$\mathcal{A}=A$ gives rise to an
effective action
\begin{equation}
S_{\rm eff} = \frac{1}{2D}\left[ \frac{1}{2} \int_0^{t_f} \left(\dot{x}+\mu\right)^2\,dt + \lambda
\left( \int_0^{t_f} [x(t)]^n dt -A\right)\right]\, ,
\label{mu_action_eff.1}
\end{equation}
where $\lambda$ is the Lagrange multiplier. Again, we assume \textit{a priori} that there is a regime where
$S_{\rm eff}$ is large and hence $P_n(A|x_0)$ can then be estimated by the saddle point method. This again
means minimizing $S_{\rm eff}$ with respect to (i) all trajectories starting at $x_0$ at $t=0$ and ending at $x=0$ at $t=t_f$
while staying positive in between, (ii) all possible values of $t_f$, and (iii)
all $\lambda$ so as to impose the constraint $\mathcal{A}=A$. Once the optimal path is found, the distribution
$P_n(A|x_0)$ can be evaluated from the optimal action
\begin{equation}
-\ln P_n(A|x_0)\simeq S_{\rm opt}= S_{\rm eff}\Big|_{\rm optimal\, path}
\label{popt.1}
\end{equation}

The presence of the $\mu$-term alters neither the Euler-Langrange equation,
\begin{equation}\label{EL0}
\ddot{x}(t)-\lambda n x^{n-1}(t)=0,
\end{equation}
nor the energy integral
\begin{equation}\label{energy}
\frac{\dot{x}^2}{2}-\lambda x^n =E = \text{const}.
\end{equation}
For zero drift, the energy $E$ was zero. For $\mu>0$,
it will be nonzero as we will see shortly. As in the driftless case, $E$ is determined from minimizing $S_{\rm eff}$ in Eq.~(\ref{mu_action_eff.1}) with respect to $t_f$ for
fixed $x_0$ and $\lambda$. This minimization gives a condition that we will use shortly:
\begin{equation}\label{eqtf}
\frac{1}{2}\left(\dot{x}+\mu\right)^2 \big|_{t=t_f}-\lambda x^n(t_f) = 0.
\end{equation}
Before proceeding further, let us remark that for $\lambda=0$ the process is unconstrained, and the optimal path -- the ballistic trajectory
\begin{equation}\label{noiselesspath}
x(t)=x_0-\mu t
\end{equation}
-- is unaffected by the noise. In this case
\begin{equation}\label{barA}
\mathcal{A} = \int_0^{x_0/\mu} (x_0-\mu t)^n \,dt = \frac{x_0^{n+1}}{\mu\, (n+1)}=
\bar{A}\,,\quad\quad n>-1\,,
\end{equation}
where $\bar A$ is defined in Eq. (\ref{abar_def}.
This is thus the mean value of $\mathcal{A}$ in the limit of
$\text{Pe}\to \infty$, as we also obtained from the exact result~(\ref{sol.2}).  Let us first consider the case
$n>0$, where Eqs.~(\ref{energy}) and~(\ref{eqtf})
suffice to determine the energy $E$ of the effective Newtonian particle.

\subsubsection{$n>0$}
\label{OFMnpositive}

In this case $x^n(t_f)=0$, and it follows from Eq. (\ref{eqtf}) that
\begin{equation}
\dot x \big|_{t=t_f}= -\mu \, .
\label{eqtf2}
\end{equation}
Then, using Eq. (\ref{energy}) at $t=t_f$, we obtain $E=\mu^2/2$. As a result,
\begin{equation}\label{mu_energy}
\frac{\dot{x}^2}{2}-\lambda x^n (t) =\frac{\mu^2}{2} \quad \quad {\rm for} \,\, n>0 \quad \text{and} \quad 0\leq t\leq t_f.
\end{equation}
Once the energy is fixed, the optimal path $x(t)$ and the optimal first-passage time $t_f$ can be determined by integrating the first-order ODE (\ref{mu_energy}) with the boundary conditions $x(0)=x_0$ and $x(t_f)=0$, while $\lambda$ is set up by the constraint $\mathcal{A}=A$.  For a general $n>0$ the optimal path cannot be expressed in explicit form. However, it is possible to evaluate the optimal action $S_{\rm opt}$
as a function of $A$ in a parametric form, where the Lagrange multiplier plays the role of the parameter.

Let us first express the effective action in Eq. (\ref{mu_action_eff.1}) for the optimal path as a function of $A$ and $t_f$. Expanding $(\dot x+\mu)^2$ and using Eq.~(\ref{mu_energy}) and the condition $x(t_f)=0$,
we obtain
\begin{equation}
S_{\rm opt} =  \frac{1}{2D}\left(-\mu\, x_0+ \mu^2\, t_f + \lambda\, A\right)\, .
\label{mu_action_opt.1}
\end{equation}
When $\mu=0$ Eq.~(\ref{mu_action_opt.1}) reduces to Eq. (\ref{S_opt.1}) for the driftless case.
Our goal now is to express $A$ and the  optimal value of  $t_f$ in
terms of $\lambda$ and the parameters $D$, $\mu$ and $x_0$. The nature of the
optimal trajectory depends on the parameter $\lambda$. Consider first $\lambda>0$,
where the effective potential $V(x)=-\lambda x^n$ is negative for all $x>0$.
Here the effective Newtonian particle with fixed energy $E=\mu^2/2$,
that starts at $x_0>0$ can reach $x=0$ only if it moves monotonically
toward $0$. The situation is different for $\lambda<0$. Note that
the $\lambda$ can not be arbitrarily negative,
since that the particle energy $E=\mu^2/2$ cannot be smaller that its potential energy. Using Eq. (\ref{mu_energy}) at $t=0$, we see that $\lambda$ cannot be smaller than
$-\lambda_c$, where
\begin{equation}
\lambda_c = \frac{\mu^2}{2\,x_0^n}\, .
\label{lc_def}
\end{equation}

\begin{figure}[h]
\includegraphics[width=0.35\textwidth,clip=]{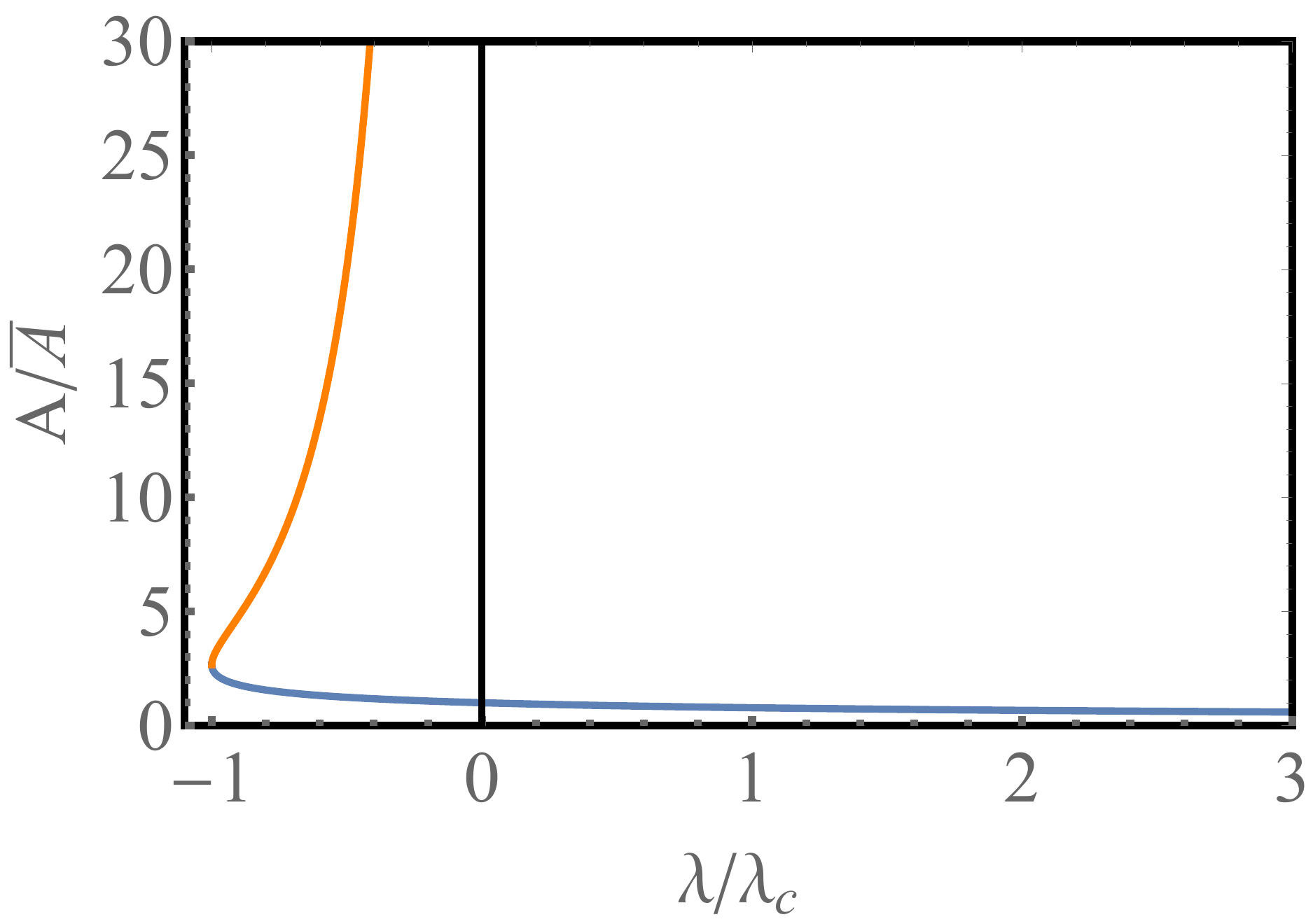}
\caption{$A/\bar{A}$ versus $\lambda/\lambda_c$, as described by
Eqs.~(\ref{zmu1.I}) and~(\ref{right_zy.1}) for $n=1$.
For fixed $\lambda>0$ there is only one solution for the
optimal path $x(t)$ and $A$, corresponding to the lower branch (branch 1).
For $-\lambda_c<\lambda<0$ there are two solutions,
corresponding to the lower and upper branches (branches 1 and 2), denoted
by the blue and orange lines, respectively.
}
\label{fig:Avslambda}
\end{figure}

Now consider $-\lambda_c\leq\lambda<0$. In this case, the potential $V(x)=-\lambda x^n$ is positive for $x>0$, and there are two possible solutions for $x(t)$ with the same $\lambda$: a monotone decreasing one and a non-monotone one.  For the non-monotone solution $x(t)$ first increases until it reaches the reflection point $x_m = \left(2 \mu^2/|\lambda|\right)^{1/n}$,
where ${\dot x}=0$, gets reflected and decreases to zero.
For the same $\lambda$ the non-monotone solution yields a larger value of $A$ than the monotone one. Figure~\ref{fig:Avslambda} depicts the dependence of $A$ on $\lambda$, which is described by Eqs.~(\ref{zmu1.I}) and~(\ref{right_zy.1}) below,   in the particular case of $n=1$. One can see the lower branch of the $A(\lambda)$-dependence  (branch 1),  and the upper branch (branch 2). We now compute the optimal action~(\ref{mu_action_opt.1}) separately for branches 1 and 2.

\begin{figure}
\includegraphics[width=0.30\textwidth,clip=]{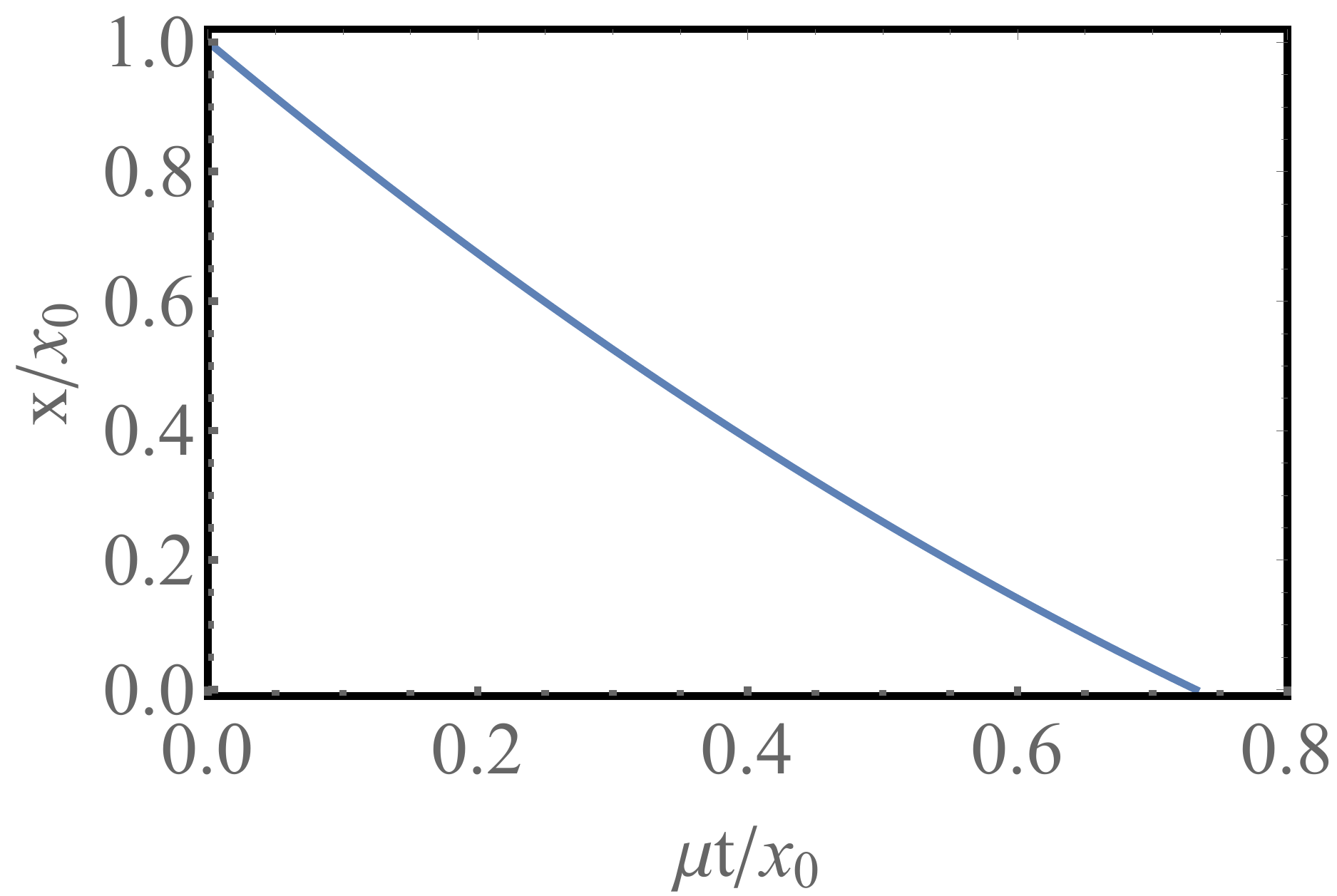}
\includegraphics[width=0.30\textwidth,clip=]{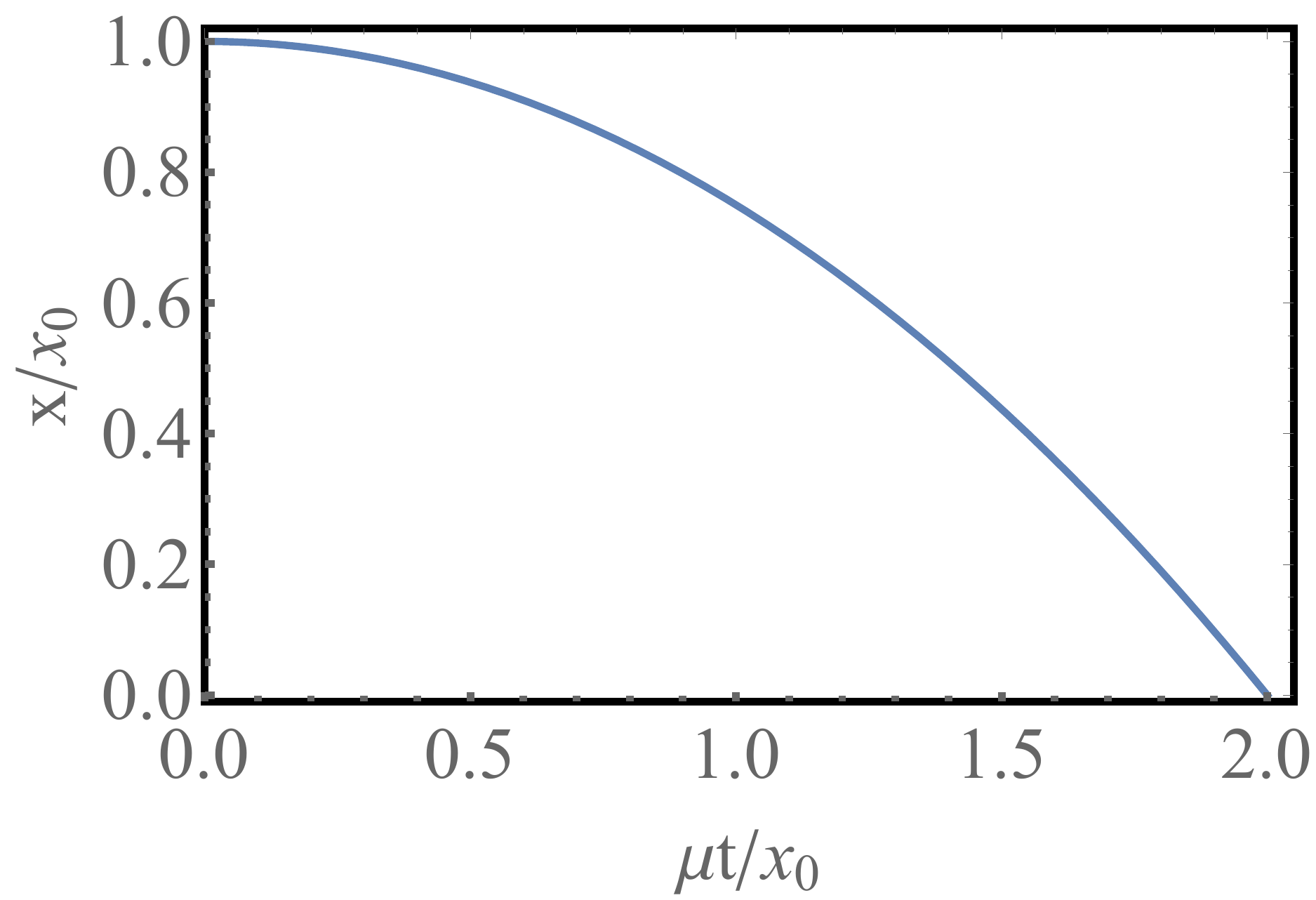}
\includegraphics[width=0.30\textwidth,clip=]{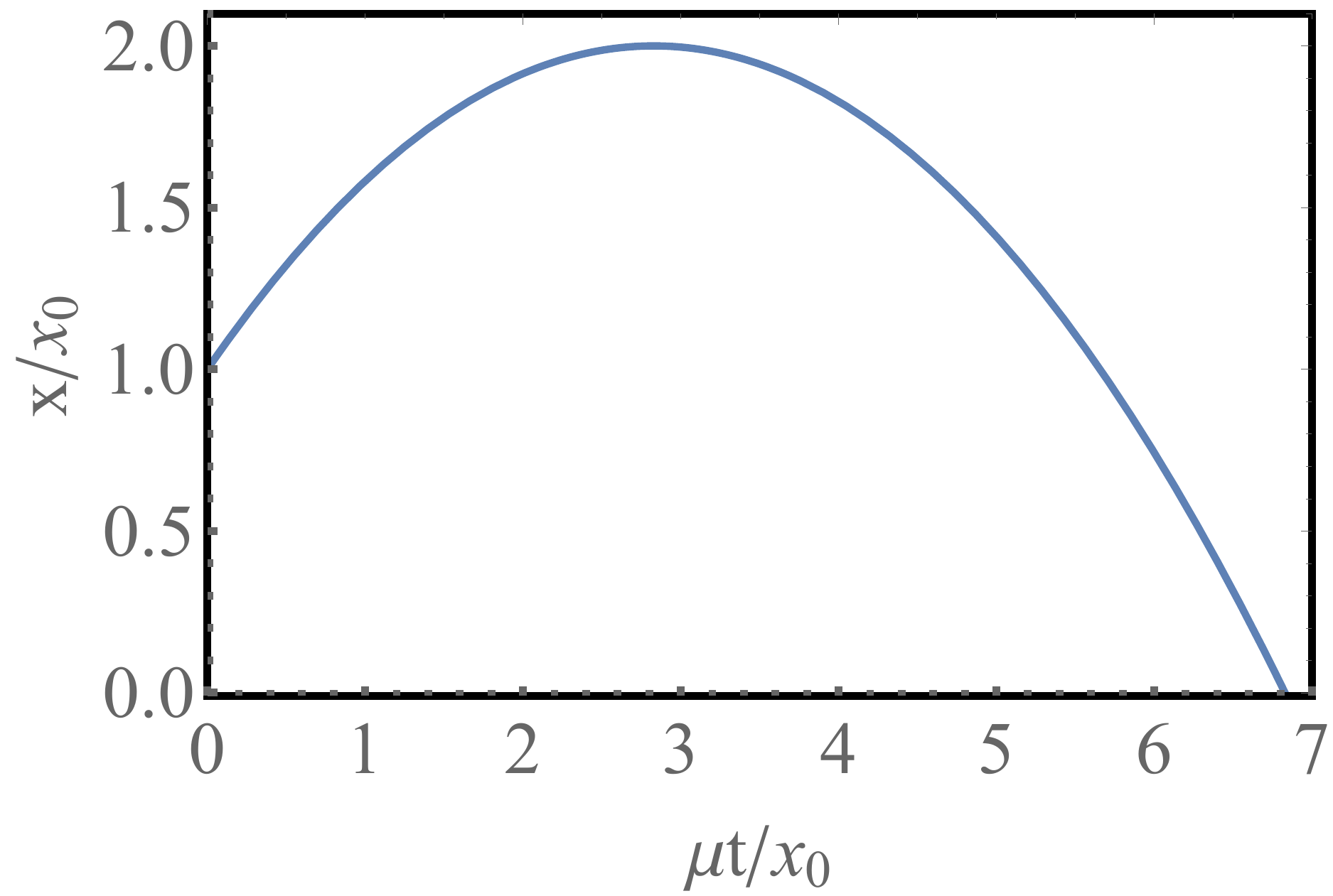}
\caption{Examples of the optimal paths $x(t)$ corresponding to the branch 1 (the left panel) and branch 2 (the right panel), see the main text. At $\lambda=-\lambda_c$ (the middle panel) $\dot x$  vanishes at $t=0$. In these examples $n=1$, so that $x(t)$ is a parabola.}
\label{threeplots}
\end{figure}

\vskip 0.4cm

\noindent {\bf Branch 1: $-\lambda_c<\lambda<+\infty$.} Here we have only monotone trajectories
with $\dot x<0$, see the left panel of Fig.~\ref{threeplots}. Hence, from
Eq. (\ref{mu_energy}), taking the negative root, we have
\begin{equation}
\frac{dx}{dt}= -\sqrt{2\,\lambda\, x^n+ \mu^2}\,.
\label{dotx_I}
\end{equation}
Integrating this ODE from $t=0$  to $t=t_f$ subject to $x(0)=x_0$ and $x(t_f)=0$, we obtain
\begin{equation}
t_f= \int_0^{x_0}\frac{dx}{\sqrt{2\lambda x^n+\mu^2}}\,.
\label{tfmu1.I}
\end{equation}
It is convenient to define
the dimensionless parameter
\begin{equation}
a= \frac{\lambda}{\lambda_c}\, .
\label{a_def}
\end{equation}
For branch 1 we have $-1<a<+\infty$. Rescaling $x= x_0\, u$ in Eq.~(\ref{tfmu1.I}),
we obtain
\begin{equation}
t_f = \frac{x_0}{\mu} \int_0^1 \frac{du}{\sqrt{1+ a\, u^n}}=
\frac{x_0}{\mu}\,\, {}_2F_1(1/2,1/n,1+1/n,-a)\, ; \quad -1<a<\infty,\quad n>0
\label{tfmu2.I}
\end{equation}
where ${}_2F_1(a,b,c,z)$ is the hypergeometric function~\cite{GR}. Now we express $A$ via $a$ from Eq.~(\ref{a_def}):
\begin{equation}
A= \int_0^{t_f} x^n(t)\, dt= \int_0^{x_0} x^n(t)\, \Big|\frac{dt}{dx}\Big|\,dx=
\int_0^{x_0} \frac{x^n\, dx}{\sqrt{2\lambda x^n+\mu^2}}\, .
\label{Amu1_I}
\end{equation}
This can be recast as
\begin{equation}
A= \frac{x_0^{n+1}}{\mu} \int_0^{1} \frac{u^n\, du}{\sqrt{1+a\, u^n}}= \frac{x_0^{n+1}}{\mu\, (n+1)}\,\,
{}_2F_1(1/2,1+1/n,2+1/n,-a)\, ; \quad -1<a<\infty\,.
\label{Amu2.I}
\end{equation}
For $a=0$ Eqs.~(\ref{tfmu2.I}) and (\ref{Amu2.I}) yield the unconstrained (noiseless) values $t_f=x_0/\mu$, and $A=\bar{A}$ from Eq.~(\ref{barA}), respectively.
Introducing the dimensionless variable $z=A/{\bar A}$, we rewrite Eq.~(\ref{Amu2.I}) as
\begin{equation}
z= \frac{A}{\bar A}= {}_2F_1(1/2,1+1/n,2+1/n,-a)\, .
\label{zmu1.I}
\end{equation}
The limiting value $a=-1$ corresponds to
\begin{equation}
z_c = z(a=-1)= {}_2F_1(1/2,1+1/n,2+1/n,1)= \sqrt{\pi}\, \frac{\Gamma(2+1/n)}{\Gamma(3/2+1/n)}\, .
\label{zc.I}
\end{equation}
Hence branch 1 is valid for $0<z\le z_c$. Plugging the expressions for $t_f$ from Eq. (\ref{tfmu2.I}) and $z=A/\bar{A}$  from Eq. (\ref{zmu1.I}) into Eq.~(\ref{mu_action_opt.1}), we see that $S_{\rm opt}$ can be written
in the scaling form
\begin{equation}
S_{\rm opt}= \text{Pe}\, \Phi_n^{(1)}\left(\frac{A}{\bar A}\right)\, ,
\label{Sopt1.I}
\end{equation}
where the scaling function for branch 1, $\Phi_n^{(1)}(z)$, is given in a parametric form by the equations
\begin{eqnarray}
z &=& {}_2F_1(1/2,1+1/n,2+1/n,-a)\,, \nonumber \\
\Phi_n^{(1)} &= & -1 + {}_2F_1(1/2,1/n,1+1/n,-a) + \frac{a}{2(n+1)}\, z(a)\, ,
\label{Phinz.I}
\end{eqnarray}
with the parameter $-1\leq a<\infty$. The limiting behaviors of $\Phi_n^{(1)}(z)$ as $z\to 0$
(that is, $a\to \infty$) and $z\to 1$ (that is, $a\to 0$)
are the following:
\begin{eqnarray}
\Phi_n^{(1)}(z) \simeq \begin{cases}
& \frac{2(n+1)}{(n+2)^2 z}\,, \quad \quad \quad \quad \quad \;\;\;z\to 0\,. \\
\\
& \frac{(2n+1)}{2\,(n+1)^2}\, (z-1)^2 \,,\quad \quad z\to 1\,.
\end{cases}
\label{Phinz_asymp.I}
\end{eqnarray}
Plugging the $z\to 0$ asymptotic from Eq.~(\ref{Phinz_asymp.I}) in Eq. (\ref{Sopt1.I}) and using Eq. (\ref{popt.1})
leads to the first line of Eq. (\ref{asymp.0}) of the Introduction; it is independent of the drift $\mu$.
In its turn, the $z\to 1$ asymptotic in Eq.~(\ref{Phinz_asymp.I}) describes a Gaussian behavior of
$P_n(A|x_0)$ near its mean value $A=\bar A$:
\begin{equation}
-\ln P_n(A|x_0)= S_{\rm opt}\left(|A-\bar{A}|\ll \bar{A}\right) \simeq \frac{(2 n+1)\mu^3}{4D x_0^{2 n+1}}
\left(A-\bar{A}\right)^2 \, ,
\label{gaussian_I}
\end{equation}
with the  variance $\sigma^2= 2\,D\,x_0^{2n+1}/{(2n+1)\mu^3}$. Notice that $\sigma/\bar{A}$ scales as $1/\sqrt{\text{Pe}}$, as to be expected for small Gaussian fluctuations.

\vskip 0.4cm

\noindent {\bf Branch 2: $-\lambda_c<\lambda<0$.} Here $t_f$ and $A$ come from two trajectory segments [see the right panel
of Fig.~\ref{threeplots}], which are governed by the equations
\begin{eqnarray}
\frac{dx}{dt} &= & \sqrt{2\, \lambda \, x^n + \mu^2}\, , \,\,\,\quad 0<t<t_m \,,\label{upward} \\
\frac{dx}{dt} &= & -\sqrt{2\, \lambda \, x^n + \mu^2}\, , \quad 0<t<t_m\,,\label{downward}
\end{eqnarray}
where $t_m$ is the reflection time of the Newtonian particle from
$x=x_m=\left(2 \mu^2/|\lambda|\right)^{1/n}$.  We obtain
\begin{equation}
t_f= \int_{x_0}^{x_m} \frac{dx}{\sqrt{2\lambda x^n+\mu^2}}+ \int_0^{x_m} \frac{dx}{\sqrt{2\lambda x^n+\mu^2}}\, .
\label{tfmu1.II}
\end{equation}
Using the same notation and rescalings as for the branch 1 (but  $-1<a<0$ now), we can rewrite Eq.~(\ref{tfmu1.II}) as
\begin{equation}
t_f= \frac{x_0}{\mu} \left[ \int_1^{(-a)^{-1/n}} \frac{du}{\sqrt{1+a\,u^n}}+ \int_0^{(-a)^{-1/n}}
\frac{du}{\sqrt{1+a\,u^n}}\right]\, = \frac{x_0}{\mu} \frac{2 B(1/n,1/2)-
B_{-a}(1/n,1/2)}{n (-a)^{1/n}}\, ,
\label{tfmu3.II}
\end{equation}
where $B_x[a,b]= \int_0^x y^{a-1}\,(1-y)^{b-1}\,dy$ is the incomplete beta function (with $x\le 1$) and
$B[a,b]= B_1[a,b]$ is the standard beta function~\cite{GR}.

The calculation of $A$ is very similar, therefore we give only the
final result for it. Similarly to the branch 1,  the rescaled quantity $z=A/{\bar A}$ can be expressed as
\begin{equation}
z= \frac{A}{\bar A}=  \frac{n+1}{n} (-a)^{-\frac{n+1}{n}}\, \left[
\frac{2\,\sqrt{\pi}\,\Gamma(1+1/n)}{\Gamma(3/2+1/n)}-B_{-a}(1+1/n,1/2)
\right]\, , \quad \quad -1<a<0\,.
\label{right_zy.1}
\end{equation}
When $a$ approaches its minimum value $a=-1$, $z$ approaches $z_c$ [given by Eq.~(\ref{zc.I})] above, therefore the branch 2 is valid for $z\ge z_c$. Substituting $t_f$ and $A= z\, \bar A$ into~Eq. (\ref{mu_action_opt.1}), we obtain
\begin{equation}
S_{\rm opt}= \text{Pe}\, \Phi_n^{(2)}\left(\frac{A}{\bar A}\right)\, ,
\label{Sopt1.II}
\end{equation}
where $\Phi_n^{(2)}(z)$ is defined parametrically by the equation
\begin{equation}
\Phi_n^{(2)}= -1 + \frac{1}{n}(-a)^{-1/n}\left\{
\frac{2\,\sqrt{\pi}\,\Gamma(1/n)}{\Gamma(1/2+1/n)}- B_{-a}[1/n,1/2]\right\}+
\frac{a}{2(n+1)}\, z(a)
\label{Phinz.II}
\end{equation}
and Eq. (\ref{right_zy.1}). Although
the solutions for the rate function $\Phi_n(z)$ for $z<z_c$ and $z>z_c$
come from two different branches, the function
$\Phi_n(z)$ is analytic at $z=z_c$ for all $n>0$.

The $z\to \infty$ asymptotic of $\Phi_n^{(2)}(z)$ is achieved in the limit of $a\to 0$, and we obtain
\begin{equation}
\Phi_n^{(2)}(z\to \infty)=\left(2^{-1/n} \, z_c\right)^{n/(n+1)}\, z^{1/(n+1)}\, .
\label{Phinz_asymp.II}
\end{equation}
Using this result in Eq. (\ref{Sopt1.I}), we obtain from Eq. (\ref{popt.1})
the asymptotic result, presented in the second line of Eq. (\ref{asymp.0}) in the Introduction.

\begin{figure}
\includegraphics[width=0.35\textwidth,clip=]{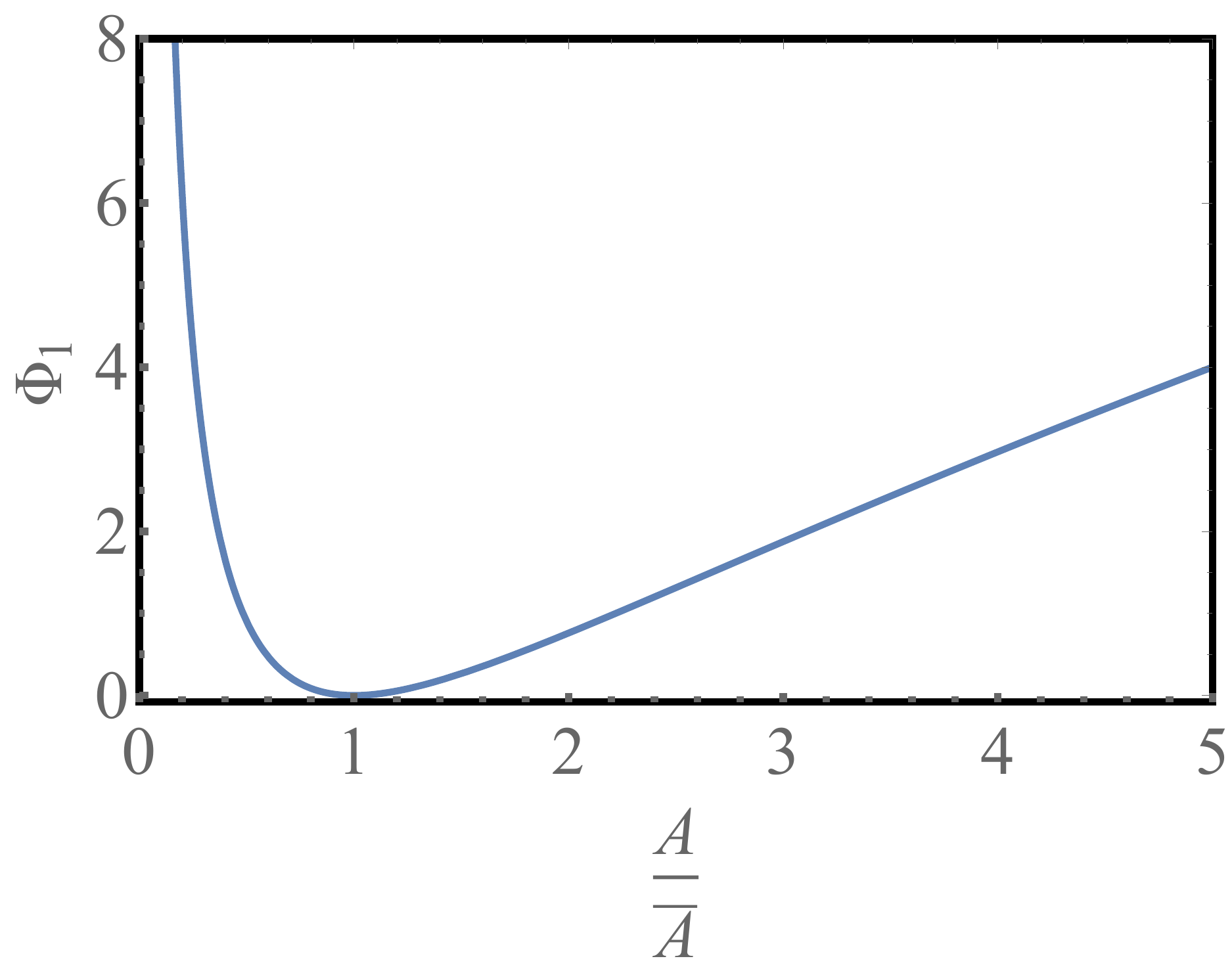}
\caption{The rate function $\Phi_1(A/\bar{A})$, described by Eq.~(\ref{ratef}).}
\label{ldffig1}
\end{figure}

For some values of $n$ the special functions
in Eqs.~(\ref{right_zy.1}) and~(\ref{Phinz.II}) become
elementary functions. A simple and instructive case
is $n=1$. Here Eqs.~(\ref{right_zy.1}) and~(\ref{Phinz.II}) become
\begin{eqnarray}
\label{n1}
  z &=& \frac{(8-4a) \sqrt{a+1}+8}{3 a^2}\,, \nonumber \\
  \Phi_1^{(2)}&=& -\frac{(a+4) \sqrt{a+1}+3 a+4}{3 a}\,.
\end{eqnarray}
Eliminating $a$, one can obtain the explicit rate function
\begin{equation}
\label{ratef}
\Phi_1(z)=\frac{2}{9z}\,\left[(1+3z)^{3/2}-9z+1\right]\,,
\end{equation}
which in fact holds for all $0<z<\infty$.  Figure~\ref{ldffig1} shows the plot of $\Phi_1(z)$.

\subsubsection{$-1<n<0$. Dynamical phase transition}
\label{min10first}

For $-1<n<0$ the OFM predicts a dramatic difference between the regimes of $A<\bar{A}$ and $A>\bar{A}$. Remarkably, for $A>\bar{A}$
the rate function $\Phi_n(z)$ is equal to zero. Indeed, to achieve an arbitrary large $A$, the particle can follow the zero-action
\emph{noiseless} path (\ref{noiselesspath}) almost until $t_f=x_0/\mu$. Arbitrarily close to $t_f$, when $x(t)$ is already very close
to zero, we can change the path a little, and make the functional (\ref{funcxn.1}) arbitrary large. The resulting action can be made
arbitrary small.  For $A<\bar{A}$ the action is nonzero, and we will calculate it shortly. These calculations will show that the
system exhibits a dynamical phase transition\footnote{A sharp transition occurs only in the limit of $\text{Pe} \to \infty$. We
expect that at finite but large $\text{Pe}$, the transition is smoothed on a narrow interval around $A=\bar{A}$, the width of
which scales as a negative power of $\text{Pe}$.} at $A=\bar{A}$.

\begin{figure}[h]
\includegraphics[width=0.35\textwidth,clip=]{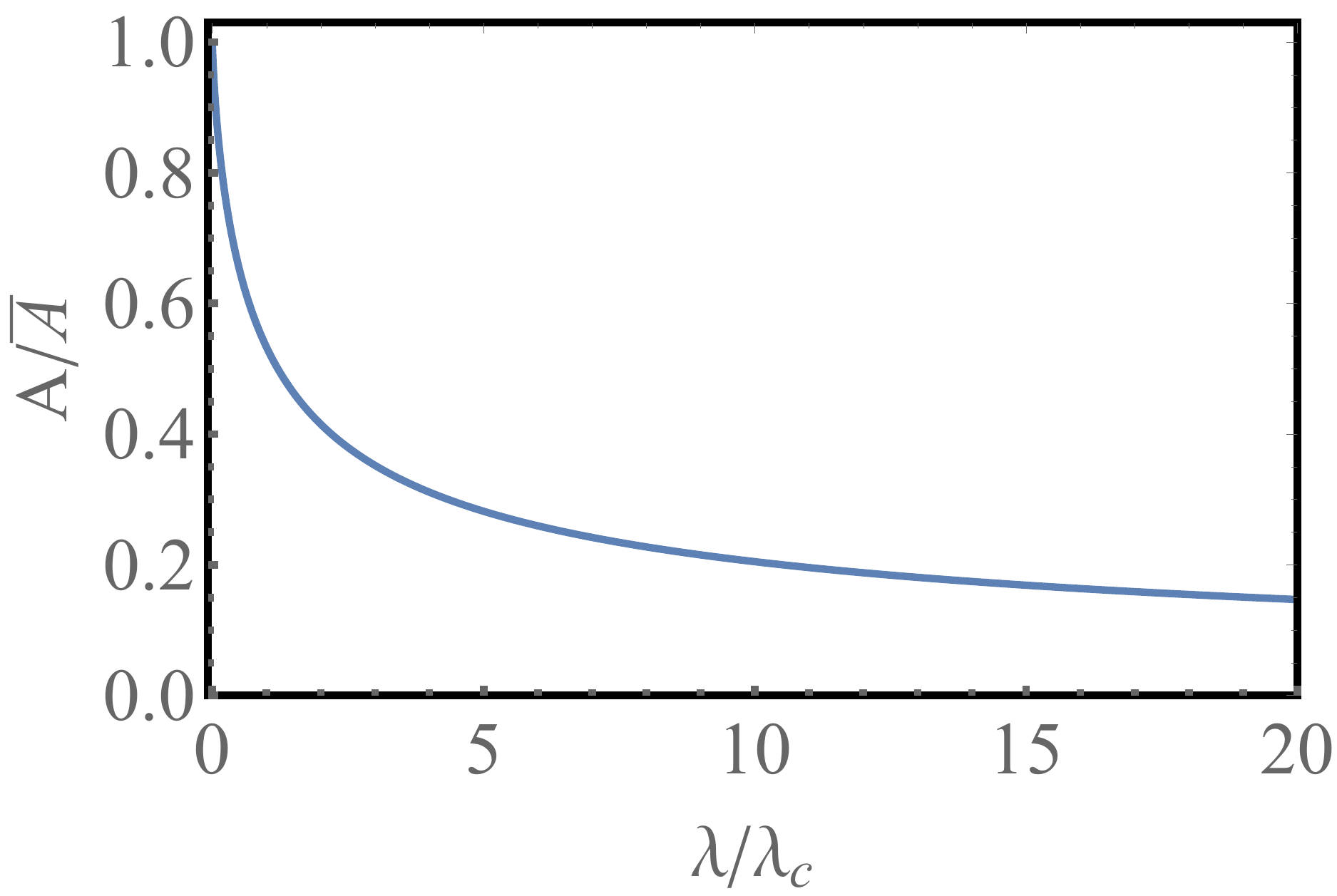}
\caption{$z=A/\bar{A}$ versus $a=\lambda/\lambda_c$, where $\lambda_c=\mu^2/(2x_0^n)$,
for $-1<n<0$ and $A<\bar{A}$ as described by Eq.~(\ref{zmu1.IL}) . Here for any $\lambda>0$ there is only one solution for the optimal path $x(t)$ and $A$. The specific example is $n=-2/3$, where
$z=\sqrt{1+a}-\sqrt{a}$.}
\label{Avslambdatwothirds}
\end{figure}

Let us determine the optimal path and the action for $A<\bar{A}$, and start with determining the
energy $E$ of the effective Newtonian particle. For $-1<n<0$,  $x^n(t_f)$ diverges. Therefore,
instead of Eq.~(\ref{eqtf}),  we will use a different argument. Let us express $\mathcal{A}$ in terms of $E$ and $\lambda$.
For $n<0$ and $A<\bar{A}$, the optimal path $x(t)$ must be monotone decreasing, so
the energy integral (\ref{energy}) yields Eq.~(\ref{dotx_I}). As a result,
\begin{equation}
\label{AvsE}
  \mathcal{A} = \int_0^{t_f} [x(t)]^n\,dt = \int_0^{x_0}\frac{dx\,x^n}{\sqrt{2(E+\lambda x^n)}}\,.  \\
\end{equation}
At $\lambda=0$ the optimal path is noiseless, and Eq.~(\ref{AvsE}) must give
$\mathcal{A}=\bar{A}$, as in Eq.~(\ref{barA}). This leads to $E=\mu^2/2$ as in the case of $n>0$.
Equation~(\ref{mu_action_opt.1}) remains valid here, and we need to express
$A$ and the  optimal value of  $t_f$ through  $\lambda$, $D$, $\mu$ and $x_0$.
Since $A<\bar{A}$, $\lambda$ must be positive [see Eq.~(\ref{AvsE})].
Figure \ref{Avslambdatwothirds} shows the dependence of $A/\bar{A}$ on
$a=\lambda/\lambda_c$, which is described by Eq.~(\ref{zmu1.IL}),  in the particular case $n=-2/3$.

To reach $x=0$ the particle must move toward $x=0$ from the start, and we again arrive at
Eqs.~(\ref{dotx_I}) and~(\ref{tfmu1.I}), although for $n<0$ the final expressions are different.  Introducing $u=x/x_0$ and $a=2\lambda x_0^n/\mu^2$, we obtain
\begin{equation}
t_f = \frac{x_0}{\mu} \left[\,
   _2F_1\left(\frac{1}{2},\frac{1}{n};1+\frac{1}{n};-a\right)-\frac{a^{-1/n}
   \Gamma \left(\frac{1}{2}-\frac{1}{n}\right) \Gamma
   \left(1+\frac{1}{n}\right)}{\sqrt{\pi }}\right]\, ; \quad 0\leq a<\infty,\quad -1<n<0.
\label{tfmu2.IL}
\end{equation}
Now we express $A$ via $a$ from Eq.~(\ref{AvsE}) with $E=\mu^2/2$:
\begin{equation}
A= \int_0^{t_f} x^n(t)\, dt= \int_0^{x_0} \frac{x^n\, dx}{\sqrt{2\lambda x^n+\mu^2}}\, .
\label{Amu1_IL}
\end{equation}
In terms of the dimensionless variable $z=A/{\bar A}$, Eq.~(\ref{Amu1_IL}) gives
\begin{equation}
z= \frac{A}{\bar A}= \frac{2(n+1)}{(n+2)\sqrt{a}}\,{}_2F_1\left(\frac{1}{2},-\frac{n+2}{2
  n};\frac{1}{2}-\frac{1}{n};-\frac{1}{a}\right)\,, \quad 0<z<1\,.
\label{zmu1.IL}
\end{equation}
Plugging the expressions for $t_f$ from Eq. (\ref{tfmu2.IL}) and
$z=A/\bar{A}$  from Eq. (\ref{zmu1.IL}) into Eq.~(\ref{mu_action_opt.1}), we can represent $S_{\rm opt}$ (and
hence $-\ln P_n(A|x_0)$)
in the scaling form
\begin{equation}
-\ln P_n(A|x_0)\simeq S_{\rm opt}= \text{Pe}\, \Phi_{-1<n<0}\left(z= \frac{A}{\bar A}\right)\, ,
\label{Sopt1.IL}
\end{equation}
where the scaling function $\Phi_{-1<n<0}(z)$ is given in a parametric form by Eq.~(\ref{zmu1.IL}) and the equation
\begin{equation}
\Phi_{-1<n<0} =  -1-\frac{a^{-1/n} \Gamma \left(\frac{1}{2}-\frac{1}{n}\right) \Gamma
   \left(1+\frac{1}{n}\right)}{\sqrt{\pi }}+\,
   _2F_1\left(\frac{1}{2},\frac{1}{n};1+\frac{1}{n};-a\right)+\frac{a z(a)}{2
   (n+1)}\, ,
\label{Phinz.IL}
\end{equation}
where $0\leq a<\infty$. Let us recall that, $z=A/{\bar{A}} \to 0$ corresponds to $a=2\lambda x_0^n/\mu^2 \to \infty $, while
$z\to 1$ from below implies $a\to 0$, as in Fig. (\ref{Avslambdatwothirds}). As $A$ increases from $0$ to $\bar{A}$, $\lambda$ decreases monotonically
from $\infty$ to $0$ for all $-1<n<0$.

What happens at $A>\bar{A}$? We note that $\lambda$ cannot be negative for $-1<n<0$: otherwise,
the effective potential energy $-\lambda x^n$ would go to plus infinity at $x\to 0$, making the arrival of our finite-energy Newtonian particle at $x=0$ impossible. Therefore, as $A$ increases beyond $\bar{A}$, $\lambda$ must stick to its value at $A=\bar{A}$, which is zero.  Since $\lambda=0$ corresponds to a noiseless classical
path as in Eq. (\ref{noiselesspath}), we obtain $S_{\rm opt}=0$, and the rate function vanishes identically, as we already argued above.

Thus summarizing, for any $-1<n<0$ at large $\text{Pe}=\mu x_0/{2D}$, the distribution $P_n(A|x_0)$ exhibits the scaling form
\begin{equation}
\label{scalingform.1}
-\ln P_n(A|x_0) \simeq  \text{Pe}\, \Phi_n\left(z= \frac{A}{\bar{A}}\right)\, ,
\end{equation}
with the rate function
\begin{numcases}
{\Phi_n(z)=} \Phi^{-}_n(z)\,,
& $z\leq 1$, \label{leftphi.1}\\
0,   & $z\geq 1$\,,\label{rightphi.1}
\end{numcases}
where $\Phi^{-}_n(z)$ is nontrivial and is given parametrically in Eq. (\ref{Phinz.IL}). In the next section we will
show that $\Phi^{-}_n(z)$ vanishes as $(1-z)^{\alpha_n}$ as $z\to 1$ from below, with an $n$-dependent exponent $\alpha_n$.
For $-1/2<n<0$, we will obtain $\alpha_n=2$, while for $-1<n<-1/2$  $\alpha_n=(1-z)^{1/(n+1)}$. The non-analytic behavior of
$\Phi_n(z)$ at $z=1$ implies a dynamical phase transition, as announced in the Introduction.

The asymptotic of $\Phi_{-1<n<0}(z)$ as $z\to 0$
(that is, $a\to \infty$) coincides with that given by the first line in Eq.~(\ref{Phinz_asymp.I}).

\begin{figure}[h]
\includegraphics[width=0.35\textwidth,clip=]{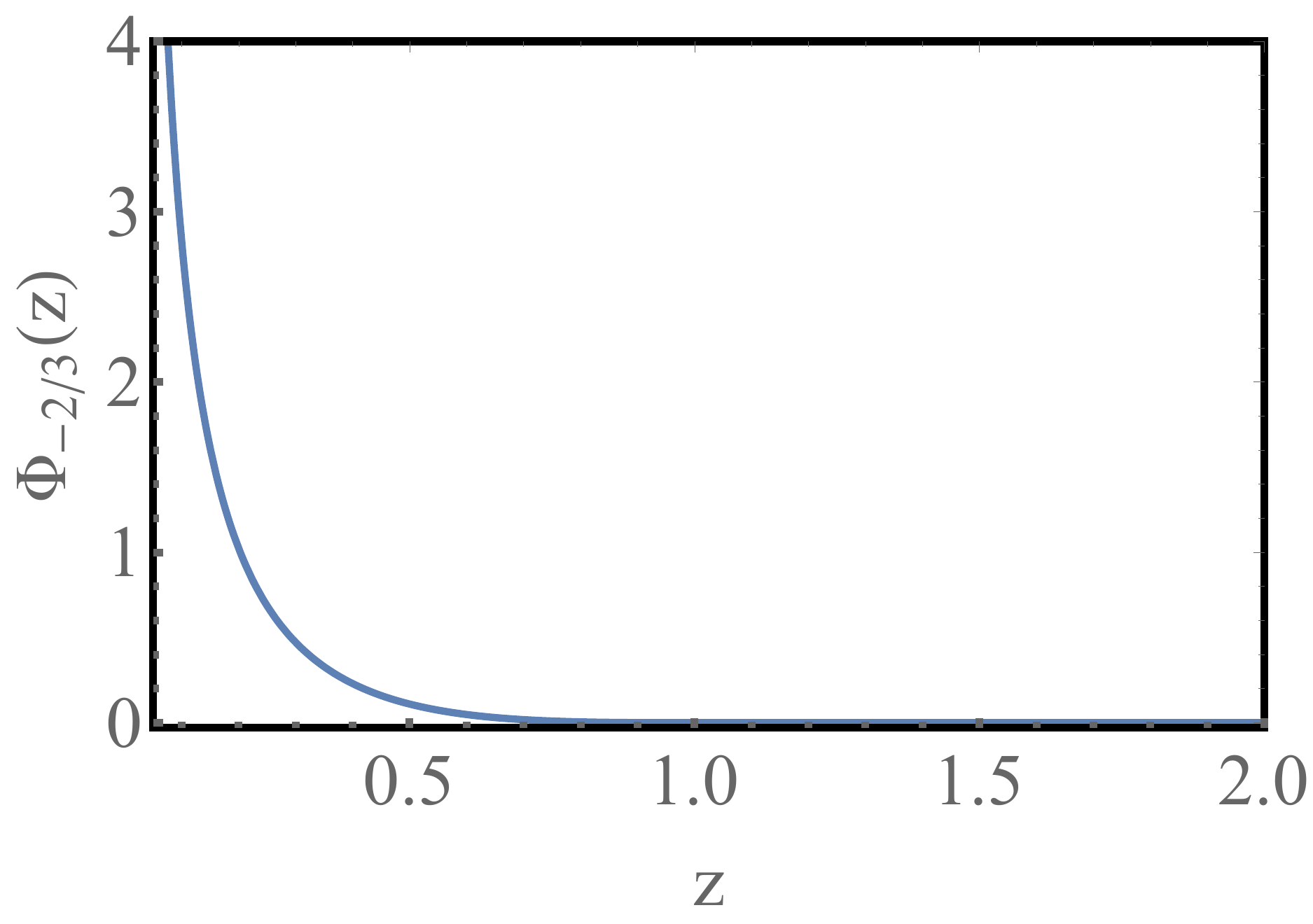}
\caption{The rate function $\Phi_{-2/3}(z)$, as in Eqs.~(\ref{nonzero}) and (\ref{zero}), for the Brownian functional $\mathcal{A}=\int_0^{t_f} [x(t)]^{-2/3} dt$. This system exhibits a dynamical phase transition of third order.}
\label{twothirds}
\end{figure}

As before, it is instructive to
consider specific values of $n$ for which the hypergeometric functions in Eqs.~(\ref{zmu1.IL}) and~(\ref{Phinz.IL}) become
elementary functions. An especially simple case if $n=-2/3$, when Eqs.~(\ref{zmu1.IL}) and~(\ref{Phinz.IL})
yield $z(a)=\sqrt{1+a}-\sqrt{a}$ and $\Phi_{-2/3} (a)=(1/2)\left[a^{3/2}+(2-a)\sqrt{1+a}-2\right]$,
respectively. Eliminating $a$ and recalling that $\Phi_{-1<n<0}=0$ for $z\geq 11$, we obtain an aesthetically
beautiful elementary expression
\begin{numcases}
{\Phi_{-2/3}(z)=}
\frac{(1-z)^3 (z+3)}{8 z}\,, & $z\leq 1$\,,\label{nonzero}\\
 0,& $z\geq 1$\,, \label{zero}
\end{numcases}
depicted in Fig.~\ref{twothirds}. It describes a dynamical phase transition of the third order at $z=1$.

\section{$\Phi_n(z)$ via WKB approximation}
\label{alternative}

In this section we provide an alternative perturbative derivation of the rate function $\Phi_n(z)$ for all $n>-1$ in
the large-$\text{Pe}$ limit, starting
from the exact differential equation (\ref{mu_diff.2}). The method we use here is a variant of the dissipative WKB approximation~\cite{Orszag}. We will see that it reproduces exactly the OFM result for all $n>-1$ in the large $\text{Pe}$-limit.
Moreover, it provides a
different representation of the rate function $\Phi_n(z)$ which is somewhat easier for the asymptotic analysis near $z=1$ to
determine the order of the dynamical phase transition at $z=1$ for $-1<n<0$.

Our starting point is the exact differential equation (\ref{mu_diff.2}) satisfied by the Laplace transform $Q_p(x_0)$.
We start with the scaling ansatz (as anticipated from our OFM analysis in the previous section)
\begin{equation}
P_n(A)\sim \exp\left[- \frac{\mu x_0}{2D}\, \Phi_n\left(\frac{A}{\bar A}\right)\right]\,, \quad\,\, {\bar A}= \frac{x_0^{n+1}}{\mu\,
(n+1)} \,,
\label{scaling.10}
\end{equation}
in the Laplace transform $Q_p(x_0)= \int_0^{\infty} e^{-p\, A}\, P_n(A)\, dA$. We obtain, up to pre-exponential factors,
\begin{eqnarray}
Q_p(x_0) &\sim & \int_0^{\infty} dA\, \exp\left [-p\, A - \frac{\mu x_0}{2D}\, \Phi_n\left(\frac{A}{\bar A}\right)\right]
\nonumber \\
& \sim & \int_0^{\infty} dz\, \exp\left\{- \frac{\mu x_0}{2D}\, \left[ \Phi_n(z) +  c\, p\, x_0^n \, z\right]
\right\}\,,\;\;\quad c= \frac{2\,D}{\mu^2\, (n+1)}\, .
\label{laplace.1}
\end{eqnarray}
When the P\'{e}clet number $\text{Pe}=\mu x_0/2D$ is large, we can evaluate the integral in Eq. (\ref{laplace.1}) by the saddle point method, while keeping the product $u=c\, p\, x_0^n$
fixed. This gives us the scaling ansatz in the Laplace space, for any $n>-1$:
\begin{equation}
Q_p(x_0) \sim \exp\left[- \frac{\mu x_0}{2D}\, \Psi_n(u= c\, p\, x_0^n)\right]\,,
\label{scaling_laplace.1}
\end{equation}
where the scaling function $\Psi_n(u)$ is given by the Legendre transform
\begin{equation}
\Psi_n(u) = \min\limits_z\, \left[ u\, z+ \Phi_n(z)\right]\, .
\label{psi_def.1}
\end{equation}
Conversely, once $\Psi_n(u)$ is known, we can extract $\Phi_n(z)$ via the inverse Legendre transform
\begin{equation}
\Phi_n(z) = \max\limits_u\, \left[ -u\, z + \Psi_n(u)\right]\, .
\label{inverse_legendre.1}
\end{equation}
Our immediate task, therefore, is to determine $\Psi_n(u)$ from Eq.~(\ref{mu_diff.2}) for $Q_p(x_0)$.   Seeking the solution in the eikonal form
\begin{equation}\label{WKBansatz}
Q_p(x_0)= e^{-s_p(x_0)/2D},
\end{equation}
we obtain an exact equation for $s_p(x_0)$:
\begin{equation}\label{secondorder}
-\frac{1}{2}D s_p''(x_0)+\frac{1}{4} [s_p'(x_0)]^2+\frac{\mu}{2} s_p'(x_0) - D p x_0^n =0.
\end{equation}
The WKB approximation is again based on the large parameter $\text{Pe}\to \infty$.
In the leading WKB order \cite{Orszag} we can neglect the second-derivative term in Eq.~(\ref{secondorder}) and express $s_p'(x_0)$ via $x_0$ from the ensuing quadratic equation for $s_p'(x_0)$\footnote{When solving the quadratic equation, we should discard the solution with the minus sign,  to avoid a divergence of $Q_p(x_0)$ at $x_0\to \infty$.}:
\begin{equation}\label{spprime}
s_p'(x_0) = -\mu+\sqrt{\mu^2+4 D p x_0^n}.
\end{equation}
Integrating this first-order equation, we obtain
\begin{equation}\label{sp(x)}
s_p(x_0)= -\mu x_0+\int_0^{x_0} \sqrt{\mu^2+4 D p x_0^n}\,dx.
\end{equation}
Comparing Eqs.~(\ref{scaling_laplace.1}) and (\ref{WKBansatz}), we obtain $\Psi_n(u) = s_p(x_0)/(\mu x_0)$, where
$x_0$ should be expressed through $u=c\, p\, x_0^n$. As in the OFM, the calculations in the cases $n>0$ and $-1<n<0$ are slightly different, and we perform
them separately\footnote{For $n=0$ the WKB result for $Q_p(x_0)$, described by Eqs.~(\ref{WKBansatz}) and~(\ref{sp(x)}), is exact and coincides with Eq.~(\ref{n0_sol.1}). Here $s_p(x_0)$ is proportional to $x_0$, and the neglected term with $s_p''(x_0)$ in Eq.~(\ref{secondorder}) vanishes identically.}.

\subsection{$n\geq 0$}
\label{alternativepos}

Here Eq.~(\ref{sp(x)}) yields, after rescalings,
\begin{equation}
\Psi_n(u) = -1+ \frac{1}{n\, u^{1/n}}\, \int_0^u \sqrt{1+2\, (n+1)\, v}\,\, {v}^{\frac{1}{n}-1}\, dv =
 -1 + {}_2F_1\,[-1/2, 1/n, 1+1/n, - 2\, (n+1)\, u]\, .
\label{sol_psi.1}
\end{equation}
As a result, the rate function $\Phi_n(z)$  is given by
\begin{equation}
\Phi_n(z)= \max\limits_u\, \left\{- u\, z -1 + {}_2F_1\,[-1/2, 1/n, 1+1/n, - 2\, (n+1)\, u]\,\right\}\,, \quad n\geq 0.
\label{phinz.1}
\end{equation}
This expression is more compact than, but equivalent to, Eq.~(\ref{Phinz.I}) that we obtained by the OFM.
The  two branches, which played a prominent role in Sec. \ref{driftOFM}, appear here in the form of two different zeros of the $u$-derivative of the function inside the curly brackets in Eq.~(\ref{phinz.1}).  Finally, we included $n=0$ in the applicability domain of Eq.~(\ref{phinz.1}) because
\begin{equation}
\lim_{n \to 0} \Psi_0(u)= -1+ \sqrt{1+2\, u}\,,
\label{psi0.1}
\end{equation}
and, as one can check,
\begin{equation}
\Phi_0(z)= \max\limits_u\, \left( - u\, z -1 + \sqrt{1+2\,u}\,\right)= \frac{(z-1)^2}{2z}\, ,
\label{phi0z.1}
\end{equation}
which coincides with the large deviation function in the exponent of the exact expression~(\ref{n0_sol.2}) for $P_0(t_f|x_0)$.

\subsection{$-1<n<0$. Dynamical phase transtion}
\label{dpt}

In this case Eq.~(\ref{sp(x)}) yields
\begin{equation}
\Psi_n(u) = -1- \frac{1}{n\, u^{1/n}} \int_u^{\infty} \sqrt{1+2\, (n+1)\, v}\,\, {v}^{\frac{1}{n}-1}\, dv =-1+\frac{\sqrt{8(n+1) u} \,
   {}_2F_1\left[-\frac{1}{2},-\frac{1}{2}-\frac{1}{n};\frac{1}{2}-\frac{1}{n};-\frac{1}{2(n+1)u}\right]}{n+2},
\label{sol_psi.1less}
\end{equation}
and the rate function is
\begin{equation}
\Phi_n(z)= \max\limits_u\, \left\{- u\, z -1+\frac{\sqrt{8(n+1) u} \,
   {}_2F_1\left[-\frac{1}{2},-\frac{1}{2}-\frac{1}{n};\frac{1}{2}-\frac{1}{n};-\frac{1}{2(n+1)u}\right]}{n+2} \right\}\,, \quad -1<n<0.
\label{phinz.1less}
\end{equation}
As $u\to 0$, the leading-order asymptotic of $\Psi_n(u)$ is $u$. The corresponding leading-order asymptotic  of the  function
$-uz+\Psi_n(u)$ is $(1-z)u$; it is positive for  $z<1$ and negative for $z>1$. As a result, the function $-uz+\Psi_n(u)$ has a local maximum at some $u=u(z)>0$ only when $z<1$. For $z\geq 1$ the maximum is always achieved at $u=0$, and the maximum value is zero. Therefore, in full agreement with the OFM results of Sec.~\ref{min10first}, the rate function $\Phi_n(z)$,
as described by Eq.~(\ref{phinz.1less}), is nonzero at $z<1$ and zero for all $z\geq 1$:
\begin{numcases}
{\Phi_n(z)=} \Phi^{-}_n(z)\,,
& $z\leq 1$, \label{leftphi}\\
0,   & $z\geq 1$\,,\label{rightphi}
\end{numcases}
where the function $\Phi^-_n(z)$ is given in Eq. (\ref{phinz.1less}). One can show that Eq. (\ref{phinz.1less}) and
the OFM result, described by Eqs.~(\ref{zmu1.IL}) and~(\ref{Phinz.IL}), are exactly equivalent.

\vskip 0.4cm

\noindent{\bf Order of the dynamical phase transition.} To determine the order of the dynamical phase transition at $z=1$, we should extract the
leading-order asymptotic of $\Phi_n(z)$ at $1-z\ll 1$. This asymptotic corresponds to the $u\to 0$ asymptotic of the function $\Psi_n(u)$ from Eq.~(\ref{sol_psi.1less}) which includes the leading linear term and the first subleading nonlinear term.  The latter asymptotic  depends on whether  $1/2<n<0$ or $-1<n<-1/2$:
\begin{numcases}
{\Psi_n(u\to 0)=} u-\frac{(n+1)^2}{2 (2n+1)}\, u^2+ \dots,
& $-\frac{1}{2}<n<0$, \label{ontheright}\\
u-C_n\,u^{-\frac{1}{n}}+ \dots,   & $-1<n<-\frac{1}{2}$\,,\label{ontheleft}
\end{numcases}
where
\begin{equation}
\label{C_n}
C_n= - \frac{\Gamma \left(1+\frac{1}{n}\right) \Gamma
   \left(-\frac{1}{2}-\frac{1}{n}\right)}{2^{1+\frac{1}{n}}\sqrt{\pi} (n+1)^{\frac{1}{n}}} >0\,.
\end{equation}
In the marginal case $n=-1/2$ we obtain a quadratic subeading term with a logarithmic correction:
\begin{equation}\label{marginal}
\Psi_{-1/2}(u\to 0) =
u- \frac{1}{4} \,u^2\,\ln \left(\frac{1}{u}\right)+\dots\,.
\end{equation}
At $-1/2<n<0$ we use Eqs.~(\ref{phinz.1less}) and~(\ref{ontheright}) to obtain
\begin{equation}\label{Phinright}
\Phi^-_n(z)\simeq \frac{(2 n+1)}{2 (n+1)^2}\, (1-z)^2\,,\quad 1-z\ll 1\,.
\end{equation}
This expression describes small one-sided Gaussian fluctuations of  $A<\bar{A}$. For all $-1/2<n<-0$, the rate function $\Phi_n(z)$ is continuous together with its first derivative at $z=1$. The second derivative has a discontinuity, so the dynamical phase transition in this case is of  second order.

For $-1<n<-1/2$, we use Eqs.~(\ref{phinz.1less}) and~(\ref{ontheleft})  to obtain,
close to $z=1$,
\begin{equation}\label{Phinleft}
\Phi^-_n(1-z\ll 1)\simeq \frac{1}{2} \pi ^{-\frac{n}{2 n+2}}
    \left[\Gamma
   \left(2+\frac{1}{n}\right) \Gamma
   \left(-\frac{1}{2}-\frac{1}{n}\right)\right]^{\frac{n}{n+1}}\,(1-z)^{\frac{1}{n+1}}.
\end{equation}
In this regime small fluctuations of $A$ around the mean value $\bar{A}$ are non-Gaussian. Furthermore, the order of the phase transition at $z=1$ now continuously depends on $n$.  As $n$ varies from $-1/2$ to $-1$, the order of transition continuously increases from $2$ to infinity. In general, it is non-integer and not even rational. This intricate behavior is quite remarkable.

In the marginal case $n=-1/2$ we use Eqs.~(\ref{phinz.1less}) and~(\ref{marginal})  to obtain, to leading order
\begin{equation}\label{minushalf}
\Phi^-_{-1/2}(1-z\ll 1)\simeq
   \frac{(1-z)^2} {\ln \left(\frac{1}{1-z}\right)}\,.
\end{equation}

Finally, using the $u\to \infty$ asymptotic of Eqs.~(\ref{sol_psi.1less}),
\begin{equation}\label{psiuinfty}
\Psi_n(u\to \infty) \simeq \frac{2 \sqrt{2(n+1) u}}{n+2}\,,
\end{equation}
and Eq.~(\ref{phinz.1less}), we checked that, to leading order, the $A\to 0$ asymptotic of $-\ln P_n(A,x_0)$ is equal to $\nu^2 x_0^{1/\nu}/(DA)$, in agreement with the exact result~(\ref{final_sol.0}) obtained for $\mu=0$.

In the particular case $n=-2/3$ the calculations simplify dramatically. Here we obtain
\begin{equation}\label{23again}
\Psi_{-2/3}(u) = \frac{1}{9} \left(2 u
   \sqrt{6 u+9}-2 \sqrt{6} \,u^{3/2}+3 \sqrt{6 u+9}-9\right),
\end{equation}
and the maximization in Eq.~(\ref{phinz.1less}) yields~Eqs.~(\ref{nonzero}) and~(\ref{zero}) of Sec.~\ref{min10first}.

\section{Discussion}
\label{discussion}

In the first part of the paper, we studied the distribution of the functional ${\cal A}= \int_0^{t_f} x^n(t)\, dt$, where
$x(t)$ represents a Brownian motion with diffusion constant $D$, starting at $x_0>0$, and $t_f$ represents
the time of the first-passage to the origin. We computed the PDF $P_n(A|x_0)={\rm Prob.}({\cal A}=A)$ exactly
for all $n>-2$, when this PDF is well defined. The PDF exhibits an essentially
singularity as $A\to 0$ and a fat tail $\sim A^{-(n+3)/(n+2)}$ as $A\to \infty$.
We complemented our exact analysis by employing the optimal fluctuation method (OFM). In OFM, one
seeks the optimal path that minimizes the effective classical action. The latter yields (the minus logarithm of) the PDF in the leading order. As we showed, the OFM correctly reproduces the leading essential singular tail at $A\to 0$. The OFM, however, cannot be used for a description of the fat tail for large $A$. This is because the power-law behavior for large $A$ arises from the contributions of many competing stochastic trajectories, and there is not one single optimal path that would dominate this tail. An added value of the OFM analysis, when it applies, is a detailed prediction of the optimal path that is not readily available in the exact method. The optimal path gives an instructive visual insight into the nature of large deviations in the system. It would be interesting to observe the optimal path in experiments/numerical simulations.

In the second part of the paper we studied the PDF $P_n(A|x_0)$ of the same functional as above, but now $x(t)$ is a Brownian motion with a nonzero drift $\mu$. In this case the PDF is well defined only for $n>-1$.

For a drift toward the origin ($\mu>0$), an explicit result for the PDF $P_n(A|x_0)$ is available only for
$n=0$. It is here where the OFM  becomes an invaluable tool, and not just a complementary technique: it allows to determine
the tails of $P_n(A|x_0)$ for any $n>-1$. The OFM results can be understood in terms of the dimensionless
P\'eclet number $\text{Pe}= \mu x_0/(2D)$ that shows the relative role of the drift and diffusion. There are two important aspects of the OFM results for $P_n(A|x_0)$:

\begin{itemize}

\item For arbitrary
$\text{Pe}$, OFM correctly predicts both tails of $P_n(A|x_0)$: $A\to 0$ and $A\to \infty$.
In the former case one again finds essential singular behavior as in the driftless case. In the latter case $P_n(A|x_0)$
has a stretched exponential tail, $-\ln P_n(A|x_0)\sim A^{-1/(n+1)}$ for $n>0$.

\item For large $\text{Pe}$, OFM captures the
exact PDF $P_n(A|x_0)$ {\em for all} $A$.
In this case we showed that $-\ln P_n(A|x_0)
\simeq \text{Pe}\, \Phi_n\left(z=A/\bar A\right)$ with $\bar A= x_0^{n+1}/{\mu (n+1)}$.
We computed the rate function $\Phi_n(z)$ analytically.

\end{itemize}

We have also shown that the OFM results can be reproduced by an alternative asymptotic perturbative theory -- the dissipative WKB approximation. While the OFM can be viewed as a WKB approximation in the `real space',
the second method is analogous,  due to a Legendre transformation involved, to a WKB approximation in the `momentum space'.

One interesting conclusion of our large-$\text{Pe}$ analysis is that, for $-1<n<0$, the function $\Phi_n(z)$ is
non-analytic at $z=1$ thus describing a dynamical phase transition. Remarkably, the order of this
transition depends on $n$ -- while it is second order for $-1/2<n<0$, the order is $1/(n+1)$ for $-1<n<-1/2$.  A sharp transition, however, occurs only in the limit of $\text{Pe} \to \infty$. We expect that at finite but large $\text{Pe}$, the transition is smoothed on a narrow interval around $A=\bar{A}$, the width of
which scales as a negative power of $\text{Pe}$.

We remark that the mechanism behind the dynamical phase transition with varying order of the transition at $-1<n<-1/2$
is very different from the mechanism of similarly looking singularities in the rate function describing
the free energy in a class of multicritical matrix models~\cite{GW1980,Wadia1980,Brezin1992,DK1993,GM1994,Marino2006}
(see also Refs. \cite{MS2014} and \cite{LMS2018} for slightly different perspectives based on extreme statistics
in matrix models). In the latter case the
order of the phase transition near the so called double scaling limit can also be varied by
varying the degree of the polynomial describing the matrix potential.
In our case, however, the transition occurs in a much simpler setting of a single particle.

Finally, we employed the OFM to study the case of $\mu<0$ (drift away from the origin), see the Appendix.  We showed that, when the process is conditioned on reaching the origin, the distribution of $\mathcal{A}$ coincides, in the limit of large $\text{Pe}$, with the distribution of $\mathcal{A}$ for $\mu>0$ with the same value of $|\mu|$. In the case of $n=0$ this duality between the two settings is known to be exact, that is to hold for any $\text{Pe}$ \cite{KrRed}. It would be interesting to see whether it is also exact for $n\neq 0$.

\section{ACKNOWLEDGMENTS}
BM was supported by the Israel Science Foundation (Grant No. 807/16) and by a Chateaubriand fellowship of the French Embassy in Israel. He is very grateful to the LPTMC, Sorbonne Universit\'{e}, for hospitality.


\begin{appendix}
\section*{Appendix. Outward drift}
\label{outward}

\renewcommand{\theequation}{A\arabic{equation}}
\setcounter{equation}{0}

For outward drift, $\mu<0$, the probability that the particle ultimately reaches $x=0$, is \cite{Rednerbook}
\begin{equation}\label{probreturn}
P_0 = e^{-\frac{|\mu| x_0}{D}} \equiv  e^{-2\text{Pe}}\,.
\end{equation}
Suppose that the P\'{e}clet number
\begin{equation}\label{newPeclet}
\text{Pe}=\frac{|\mu|x_0}{2D}
\end{equation}
is much larger than $1$. Then the probability (\ref{probreturn}) of ever reaching zero is exponentially small. Still, one can ask a similar question about the probability density of the Brownian functional $\mathcal{A}$ from Eq.~(\ref{funcxn.1}) when  the process is conditioned on reaching $x=0$.  Within the framework of the OFM, this conditional probability density is equal to the ratio of the probability densities of two different optimal paths: with and without the constraint $\mathcal{A}=A$. Equivalently, the optimal constrained action is equal to the difference of the actions of the optimal paths with and without the constraint. For completeness, we first show, within the framework of the OFM, that the unconstrained action is equal to $2\text{Pe}$ in agreement with the exact result (\ref{probreturn}). In the absence of constraint on $\mathcal{A}$, the Euler-Lagrange equation ~(\ref{EL0})  becomes simply $\ddot{x}=0$. Its solutions, obeying the initial condition $x(0)=x_0$ and the condition of reaching $x=0$ at some time $t_f$, can be written as $x(t)=x_0(1-t/t_f)$. The unconstrained action  is, therefore,
\begin{equation}\label{unconstrainedaction}
S=\frac{1}{4D}\int_0^{t_f} \left(\dot{x}+\mu\right)^2\,dt=\frac{1}{4D}\int_0^{t_f} \left(-\frac{x_0}{t_f}+\mu\right)^2\,dt=\frac{1}{4D}\left(\frac{x_0^2}{t_f}+\mu^2 t_f-2\mu x_0\right)\,.
\end{equation}
Now we should minimize this expression with respect to the first-passage time $t_f$. The minimum value of $S$ is achieved at $t_f=x_0/|\mu|$: the optimal unconstrained path describes ballistic motion with the velocity equal to the \emph{minus} deterministic drift velocity. The resulting optimal unconstrained action, as obtained from Eq.~(\ref{unconstrainedaction}), is equal to $|\mu| x_0/D=2\text{Pe}$, as to be expected from Eq.~(\ref{probreturn}). Notice that the corresponding optimal unconstrained value of $\mathcal{A}$,
\begin{equation}\label{A0}
A_0=\frac{x_0^{n+1}}{|\mu| (n+1)}\,,
\end{equation}
coincides, up to the change $\mu \to |\mu|$, with $\bar{A}$ from Eq.~(\ref{abar_def}).

Now we should find the optimal path constrained by $\mathcal{A}=A$.  The Euler-Lagrange equation~(\ref{EL0}), the initial condition $x(0)=x_0$ and the constraint $\mathcal{A}=A$ do not depend on $\mu$. The $\mu$-dependence comes only from the value of the energy $E$ of the effective Newtonian particle in Eq.~(\ref{energy}). As one can show, it is equal to $E=\mu^2/2$ as before\footnote{One way to show it, valid for all $n>-1$, is to use the fact that at $\lambda=0$ one obtains the unconstrained optimal path, for which one has $A=A_0$ as in Eq.~(\ref{A0})}. The offshoot is that, for a fixed $\lambda$, the optimal path for $\mu<0$ \emph{coincides} with the optimal path for $\mu>0$ with the same $|\mu|$. As a result, $A$ as a function of the Lagrange multiplier $\lambda$ in the two problems with the same $|\mu|$ is exactly the same.

The optimal actions in these two problems, $S_{\mu}$ and $S_{-\mu}$ (where $\mu<0$), are of course different. Let us evaluate their difference. We have
\begin{equation}\label{difference}
S_{\mu}-S_{-\mu}=\frac{1}{4D}\int_0^{t_f} \left(\dot{x}+\mu\right)^2\,dt-
\frac{1}{4D}\int_0^{t_f} \left(\dot{x}-\mu\right)^2\,dt
= \frac{1}{4D}\int_0^{t_f} 2 \mu \times 2\dot{x}\,dt =\frac{|\mu| x_0}{D}=2\text{Pe}\,,\quad \mu<0\,.
\end{equation}
Therefore, $S_{\mu}=S_{-\mu}+2\text{Pe}$. After subtracting the unconditioned action -- which is also equal to $2\text{Pe}$ -- we arrive at
\begin{equation}
-\ln P_n^{(\mu<0)}(A) \simeq S_{-\mu}=\text{Pe} \,\Phi_n\left(\frac{A}{A_0}\right),
\label{scaling.unfavor}
\end{equation}
where $\text{Pe}$ is defined in Eq.~(\ref{newPeclet}), and $\Phi_n(z)$ was calculated  in Sec. \ref{OFMnpositive} for $n>0$, and  in Sec.~\ref{min10first} for $-1<n\leq 0$. That is, all our results for the rate function in the case of favorable drift (the asymptotics, the dynamical phase transition at $-1<n<0$, \textit{etc}.) also hold for the unfavorable drift with the same value of $|\mu|$, if the latter process is conditioned on reaching $x=0$. In the particular case $n=0$ this duality has been previously established (exactly, that is for any value of $\text{Pe}$)  in Ref. \cite{KrRed}.

\end{appendix}

\end{document}